\documentclass[12pt]{article}
\usepackage{amsmath,amssymb,amsthm,amsxtra,overpic,bbm,bm,epsfig,subfigure}
\usepackage{hyperref}
\usepackage{mathrsfs}
\usepackage{graphicx}
\usepackage{multirow}
\usepackage{color}
\usepackage{comment}
\usepackage{epstopdf}
\usepackage{float}
\usepackage{cite}
\usepackage{hyperref}
\usepackage{url}
\usepackage{slashed,stmaryrd}
\usepackage{longtable}
\usepackage{extarrows}

\begin{document}
\def\thefootnote{\fnsymbol{footnote}}
\vspace{0.2cm}
\begin{center}
{\Large\bf Hilbert Series for Leptonic Flavor Invariants in \\ the Minimal Seesaw Model}
\end{center}
\vspace{0.2cm}

\begin{center}
{\bf Bingrong Yu}~$^{a,~b}$~\footnote{E-mail: yubr@ihep.ac.cn}
\quad
{\bf Shun Zhou}~$^{a,~b}$~\footnote{E-mail: zhoush@ihep.ac.cn (corresponding author)}
\\
\vspace{0.2cm}
{\small
$^a$Institute of High Energy Physics, Chinese Academy of Sciences, Beijing 100049, China\\
$^b$School of Physical Sciences, University of Chinese Academy of Sciences, Beijing 100049, China}
\end{center}
	
\vspace{0.5cm}

\begin{abstract}
In this paper, we examine the leptonic flavor invariants in the minimal seesaw model (MSM), in which only two right-handed neutrino singlets are added into the Standard Model in order to accommodate tiny neutrino masses and explain cosmological matter-antimatter asymmetry via leptogenesis mechanism. For the first time, we calculate the Hilbert series (HS) for the leptonic flavor invariants in the MSM. With the HS we demonstrate that there are totally 38 basic flavor invariants, among which 18 invariants are CP-odd and the others are CP-even. Moreover, we explicitly construct these basic invariants, and any other flavor invariants in the MSM can be decomposed into the polynomials of them. Interestingly, we find that any flavor invariants in the effective theory at the low-energy scale can be expressed as rational functions of those in the full MSM at the high-energy scale. Practical applications to the phenomenological studies of the MSM, such as the sufficient and necessary conditions for CP conservation and CP asymmetries in leptogenesis, are also briefly discussed.
\end{abstract}

\newpage

\def\thefootnote{\arabic{footnote}}
\setcounter{footnote}{0}
\section{Introduction}
Neutrino oscillation experiments have firmly established that neutrinos are massive particles and the flavor mixing is significant in the leptonic sector~\cite{PDG2020,Xing2020}. This compelling evidence for neutrino masses and flavor mixing indicates that the Standard Model (SM) is actually incomplete and new physics indeed exists. In order to accommodate tiny neutrino masses in a natural way, one can extend the SM with $N$-generation right-handed (RH) neutrino singlets $\nu_{i\rm R}^{}$ (for $i=1,2,...,N$), for which the Majorana mass term is allowed. More explicitly, the ${\rm SU}(2)_{\rm L}^{}\otimes {\rm U}(1)_{\rm Y}^{}$ gauge-invariant Lagrangian responsible for lepton masses, flavor mixing and CP violation reads
\begin{eqnarray}
\label{eq:Lagrangian before SSB}
-{\mathscr L}_{\rm lepton}^{}=\overline{\ell_{\rm L}^{}}Y_l^{}H l_{\rm R}^{}+\overline{\ell_{\rm L}^{}}Y_\nu^{}\tilde{H}\nu_{\rm R}^{}+\frac{1}{2}\overline{\nu_{\rm R}^{\rm C}}M_{\rm R}^{}\nu_{\rm R}^{}+{\rm h.c.}\;,
\end{eqnarray}
where ${\ell }_{\rm L}^{}\equiv \left(\nu_{\rm L}^{},l_{\rm L}^{}\right)_{}^{\rm T}$, $\tilde{H}\equiv {\rm i}\sigma_2^{}H_{}^{*}$ with $H\equiv\left(H_{}^{+},H_{}^{0} \right)_{}^{\rm T}$ stand for the left-handed lepton doublet and the Higgs doublet, respectively. In addition, $Y_l^{}$ and $Y_\nu^{}$ are the charged-lepton and Dirac neutrino Yukawa coupling matrices while $M_{\rm R}^{}$ is the Majorana mass matrix of RH neutrinos. Note that $\nu_{\rm R}^{\rm C}\equiv {\cal C} \overline{\nu_{\rm R}^{}}^{\rm T}$ has been defined with ${\cal C}\equiv {\rm i}\gamma_{}^2\gamma_{}^0$ being the charge-conjugation matrix.

After the Higgs field acquires its vacuum expectation value $\langle H_{}^0\rangle =v/\sqrt{2}$ with $v\approx 246~{\rm GeV}$, the ${\rm SU}(2)_{\rm L}^{}\otimes {\rm U}(1)_{\rm Y}^{}$ gauge symmetry is spontaneously broken down, and the lepton mass terms and the leptonic charged-current interation are given by
\begin{eqnarray}
\label{eq:Lagrangian after SSB}
-{\mathscr L}_{\rm lepton}^{\prime}=\overline{l_{\rm L}^{}}M_l^{}l_{\rm R}^{}+\overline{\nu_{\rm L}^{}}M_{\rm D}^{}\nu_{\rm R}^{}+\frac{1}{2}\overline{\nu_{\rm R}^{\rm C}}M_{\rm R}^{}\nu_{\rm R}^{}-\frac{g}{\sqrt{2}}\overline{l_{\rm L}}\gamma_{}^{\mu}\nu_{\rm L}^{}W_{\mu}^{-}+{\rm h.c.}\;,
\end{eqnarray}
where the charged-lepton mass matrix and the Dirac neutrino mass matrix are respectively given by $M_l^{} = Y_l^{}v/\sqrt{2}$ and $M_{\rm D}^{} = Y_\nu^{}v/\sqrt{2}$, and $g$ is the coupling constant of the ${\rm SU}(2)_{\rm L}^{}$ gauge group. If the mass scale of RH neutrinos is far above the electroweak scale $\Lambda^{}_{\rm EW} = 10^{2}~{\rm GeV}$, namely, ${\cal O}(M^{}_{\rm R}) \gg \Lambda^{}_{\rm EW}$, one can integrate out heavy Majorana neutrinos and obtain the low-energy effective theory, in which the effective Majorana mass matrix of ordinary neutrinos is given by
\begin{eqnarray}
\label{eq:seesaw formula}
M_{\nu}^{}=-M_{\rm D}^{}M_{\rm R}^{-1}M_{\rm D}^{\rm T}\; .
\end{eqnarray}
In this canonical seesaw model~\cite{Minkowski1977, GellMann1979, Yanagida1980, Glashow1980, Mohapatra1980}, three light neutrinos are Majorana particles, namely, they are their own antiparticles~\cite{Majorana1937, Racah1937}. Thus the effective Lagrangian governing lepton masses, flavor mixing and CP violation at the low-energy scale becomes
\begin{eqnarray}
\label{eq:Lagrangian effective}
-{\mathscr L}_{\rm eff}^{}=\overline{l_{\rm L}^{}}M_l^{}l_{\rm R}^{}+\frac{1}{2}\overline{\nu_{\rm L}^{}}M_\nu^{}\nu_{\rm L}^{\rm C}-\frac{g}{\sqrt{2}}\overline{l_{\rm L}}\gamma_{}^{\mu}\nu_{\rm L}^{}W_{\mu}^{-}+{\rm h.c.}\;.
\end{eqnarray}
It is the mismatch between the diagonalizations of $M^{}_l$ and $M^{}_\nu$ that leads to flavor mixing and CP violation in the leptonic charged-current interaction. As one can easily verify, under the unitary transformations of the lepton fields in the flavor space, the lepton mass matrices in Eqs.~(\ref{eq:Lagrangian after SSB}) and (\ref{eq:Lagrangian effective}) will certainly change. However, physical parameters, such as the eigenvalues of lepton mass matrices corresponding to the lepton masses, are invariant under the flavor transformations. The reason is simply that lepton masses are physical observables, which should be independent of the flavor basis. The flavor invariants constructed from fermion mass matrices, which are basis-independent and contain only physical degrees of freedom, have proved to be very useful in studying the flavor structures and CP violation in the quark and leptonic sector~\cite{Jarlskog1985a, Jarlskog1985b, Wu1986, Branco1986quark, Branco1986lepton, Yu2019PLB, Yu2020ICHEP, Yu2020PRD}.

In Ref.~\cite{Wang2021}, implementing the mathematical tool of Hilbert series (HS)  and plethystic logarithm (PL) from the invariant theory~\cite{Sturmfels2008,DK2015}, we have studied the algebraic structures of the invariant ring in the low-energy effective theory with three massive Majorana neutrinos and explicitly constructed all the 34 basic flavor invariants. All the flavor invariants, built upon the matrix polynomials of $M_l^{}$ and $M_{\nu}^{}$ in Eq.~(\ref{eq:Lagrangian effective}), can be decomposed into the polynomials of these 34 basic invariants. In the canonical seesaw model with three RH neutrinos, the ungraded HS and the flavor invariants have been discussed in Refs.~\cite{JM2009,HJMT2011}. However, the HS is too complicated to be useful in the explicit construction of all the basic invariants. In the present paper, we examine the flavor invariants in the minimal seesaw model (MSM), in which the SM is extended with only two RH Majorana neutrinos~\cite{Kleppe1995, Ma1998, King1999, King2002, Lavoura2000, Frampton2002}. See, e.g., Refs.~\cite{Guo2006,XZ2020}, for recent reviews on the MSM. The main motivation for such an investigation is two-fold.
\begin{itemize}
\item The MSM is the most economical scenario where tiny neutrino masses can be generated by the seesaw mechanism and the cosmological matter-antimatter asymmetry can be explained via the leptogenesis mechanism~\cite{Fukugita1986}. Therefore, the algebraic structure of the invariant ring of the MSM is nontrivial on the one hand, and it is possible to explicitly construct all the basic flavor invariants on the other hand. This is an excellent example to explore the algebraic structure of the invariant ring in an ultraviolet (UV)-complete model.

\item Since the basic invariants in the low-energy effective theory have been found in Ref.~\cite{Wang2021}, it is interesting to examine the basic invariants in the high-energy full theory and to establish a connection between these two sets of basic invariants. For this purpose, we start with the construction of flavor invariants directly from $M_l^{}$, $M_{\rm D}^{}$ and $M_{\rm R}^{}$ in Eq.~(\ref{eq:Lagrangian after SSB}) in the full MSM. By calculating the HS and PL, we find that there are in total 38 basic invariants in the MSM, among which 18 invariants are CP-odd and the others are CP-even. Moreover, we demonstrate that any flavor invariants at the low-energy scale can be expressed as the rational functions of those at the high-energy scale.
\end{itemize}

The remaining part of this paper is structured as follows. In Sec.~\ref{sec:flavor inv} we briefly introduce the mathematical background of the invariant theory and explain the general strategy to construct the flavor invariants in a complete model, which extends the SM with $N$ generations of RH Majorana neutrinos. Then, we calculate the HS in the MSM and explicitly construct all the basic invariants in Sec.~\ref{sec:HS}.  Some further discussions are presented in Sec.~\ref{sec:discussion}, where the relationship between the flavor invariants at high- and low-energy scales is established and phenomenological applications of the flavor invariants are also discussed, including the sufficient and necessary conditions for CP conservation and the CP asymmetries in heavy Majorana neutrino decays. Finally, our main results and conclusions are summarized in Sec.~\ref{sec:summary}.

\section{Flavor Invariants}
\label{sec:flavor inv}

\subsection{Invariant Theory and Hilbert Series}
In this subsection we sketch the indispensable mathematical ingredients of the invariant theory. For a more detailed and pedagogical introduction, see, e.g., Appendix B of Ref.~\cite{Wang2021}.

Given a theory with $n$ parameters $\vec{x}=\left(x_1^{},...,x_n^{}\right)$ and a symmetry group $G$, we are interested in those quantities that are invariant under the group action, i.e.,
\begin{eqnarray}
I(\vec{x})=I\left(R(g)\vec{x}\right)\;,\quad
\forall g \in G\;,
\end{eqnarray}
where $R$ is a specific representation of $G$ and $I(\vec{x})$ is a polynomial function of $\vec{x}$. Since all the invariants are closed under the addition and multiplication, they form a ring. For a reductive group, including all the finite groups and semi-simple Lie groups, the ring is finitely generated, in the sense that all the invariants in the ring could be expressed as the polynomials of a finite number of \emph{basic} invariants. Thus these basic invariants serve as the generators of the ring.

It is worth noting that not all the basic invariants are algebraically independent, and there may exist polynomial functions of the basic invariants that are identically equal to zero~\cite{Trautner2018,Trautner2020}. These non-trivial polynomial relations between the basic invariants are known as syzygies. The maximal number of  algebraically-independent invariants is the Krull dimension of the ring and these algebraically-independent invariants are also called \emph{primary} invariants. A significant result is that the number of the primary invariants (also the Krull dimension of the ring) equals the number of the physical parameters in the theory.

In general, the number of the basic invariants, denoted as $m$, is no smaller than that of the primary invariants $r$. The special case of $m=r$ corresponds to the free ring where there is no syzygy at all. Furthermore, if the number of the syzygies equals $m-r$, then the ring is a complete intersection, otherwise a non-complete intersection.

In the invariant theory, the HS and PL provide a convenient way to count the number of basic invariants, as well as their degrees and the syzygies among them. The HS serves as the generating function of the invariants
\begin{eqnarray}
\label{eq:HS ungraded def}
{\mathscr H}\left(q\right) \equiv \sum_{k=0}^{\infty}c_k^{} q_{}^k\;,
\end{eqnarray}
where $c_k^{}$ (with $c_0^{}\equiv1$) stand for the number of linearly-independent invariants at degree $k$ while $q$ is an arbitrary complex number satisfying $\left|q\right|<1$. The HS can always be written as the ratio of two polynomial functions~\cite{DK2015}
\begin{eqnarray}
{\mathscr H}(q)=\frac{{\mathscr N(q)}}{{\mathscr D(q)}}\;,
\end{eqnarray}
where the numerator takes on the palindromic structure
\begin{eqnarray}
{\mathscr N}(q)=1+a_1^{}q+...+a_{l-1} q_{}^{l-1}+q_{}^l\;,
\end{eqnarray}
with $a_k^{}=a_{l-k}^{}$. The particular case with ${\mathscr N}(q)=1$ corresponds to the free ring. In addition, the denominator has the general form
\begin{eqnarray}
{\mathscr D}(q)=\prod_{k=1}^r (1-q^{d_k})\;,
\end{eqnarray}
encoding the information of primary invariants~\cite{Sturmfels2008, DK2015}. The total number of the factors $r$ equals the Krull dimension of the ring, or the number of the primary invariants, while the power indices $d_k^{}$ (for $k=1,2,...,r$) indicate the degree of each primary invariant.

The definition of the (ungraded) HS in Eq.~(\ref{eq:HS ungraded def}) can be generalized to the multi-graded form in a straightforward way. Suppose that there are $n$ independent building blocks to construct the invariants, then the multi-graded HS is defined as
\begin{eqnarray}
\label{eq:HS multi-graded def}
{\cal H}\left(q_1^{},...,q_n^{}\right)\equiv \sum_{k_1=0}^{\infty}...\sum_{k_n=0}^{\infty}c_{k_1...k_n}q_1^{k_1}...q_n^{k_n}\;,
\end{eqnarray}
where $q_i^{}$ (for $i=1,2,...,n$) label the degree of the $i$-th building block and satisfy $\left|q_i^{}\right|<1$, while $c_{k_1...k_n}^{}$ (with $c_{0...0}\equiv 1$) denote the number of linearly-independent invariants when the $n$ building blocks are at the degree of $(k_1^{},...,k_n^{})$, respectively. 

Given the HS, one can calculate its PL, which counts the number of the basic invariants and the syzygies. The PL of an arbitrary function $f(x_1^{},...,x_n^{})$ is defined as
\begin{eqnarray}
\label{eq:PL def}
{\rm PL}\left[f(x_1^{},...,x_n^{})\right]\equiv
\sum_{k=1}^{\infty} \frac{\mu(k)}{k}\,{\rm ln}\left[f(x_1^k,...,x_n^k)\right]\;,
\end{eqnarray}
where $\mu(k)$ is the M{\"o}bius function. It has been pointed out in Ref.~\cite{BFHH2007} that the leading positive terms of PL correspond to the basic invariants while the leading negative terms of PL correspond to the syzygies among these basic invariants. Moreover, the PL for a complete intersection ring has only a finite number of terms while for a non-complete intersection ring it is an infinite series.

Calculating HS from the definition is usually very difficult. A  systematic method to calculate HS is to make use of the Molien-Weyl (MW) formula~\cite{Molien1897,Weyl1926}
\begin{eqnarray}
\label{eq:Molien-Weyl formula}
{\cal H}\left(q_1^{},...,q_n^{}\right)=\int \left[{\rm d}\mu \right]_G {\rm PE}\left(z_1^{},...,z_{r_0}^{}; q_1^{},...,q_n^{}\right)\;,
\end{eqnarray}
where $\left[{\rm d}\mu\right]_G^{}$ denotes the Haar measure of the symmetry group $G$ and the integrand is the plethystic exponential (PE) defined as
\begin{eqnarray}
{\rm PE}\left(z_1^{},...,z_{r_0}^{};q_1^{},...,q_n^{}\right)\equiv{\rm exp}\left[\sum_{k=1}^{\infty}\sum_{i=1}^n \frac{\chi_{R_i}^{}\left(z_1^k,...,z_{r_0}^k\right)q_i^k}{k} \right]\;,
\end{eqnarray}
with $z_i^{}$ (for $i=1,2,...,r_0^{}$) the coordinates on the maximum torus of $G$, $r_0^{}$ the rank of $G$ and $\chi_{R_i}^{}$ the character function of $G$ under the representation $R_i^{}$. The MW formula reduces the computation of HS to several complex integrals, which can be performed by virtue of the residue theorem.

\subsection{Flavor Transformations and Flavor Invariants}
Then we consider the general seesaw model with $N$ generations of RH Majorana neutrinos and explain how to construct the flavor invariants in a systematic way.

The Lagrangian in the leptonic sector after the spontaneous symmetry breaking is given by Eq.~(\ref{eq:Lagrangian after SSB}) and is unchanged under the following unitary transformations in the flavor space
\begin{eqnarray}
l_{\rm L}^{}\rightarrow l_{\rm L}^{\prime}=U_{\rm L}^{}l_{\rm L}^{}\;,\quad
l_{\rm R}^{}\rightarrow l_{\rm R}^{\prime}=V_{\rm R}^{}l_{\rm R}^{}\;,\quad
\nu_{\rm L}^{}\rightarrow \nu_{\rm L}^{\prime}=U_{\rm L}^{}\nu_{\rm L}^{}\;,\quad
\nu_{\rm R}\rightarrow \nu_{{\rm R}}^{}=U_{\rm R}^{}\nu_{\rm R}^{}\;,
\end{eqnarray}
where $U_{\rm L}^{},V_{\rm R}^{}\in {\rm U}(3)$ and $U_{\rm R}^{}\in {\rm U}(N)$ are three arbitrary unitary matrices, if the lepton mass matrices transform as
\begin{eqnarray}
\label{eq:mass matrices trans}
M_l^{} \rightarrow  M_l^{\prime} = U_{\rm L}^{}M_l^{}V_{\rm R}^{\dagger}\;,\qquad M_{\rm D}^{} \rightarrow  M_{\rm D}^{\prime} = U_{\rm L}^{}M_{\rm D}^{}U_{\rm R}^{\dagger}\;, \qquad M_{\rm R}^{} \rightarrow M_{\rm R}^{\prime}=U_{\rm R}^{*}M_{\rm R}U_{\rm R}^{\dagger} \;.
\end{eqnarray}
Based on the transformation rules of the mass matrices in Eq.~(\ref{eq:mass matrices trans}) we can introduce the ``building blocks", which transform as the adjoint representations of ${\rm U}(3)$ and ${\rm U}(N)$, for constructing the flavor invariants, i.e.,\footnote{The invariants composed of other building blocks, such as $M_l^\dagger M_l^{}$, which transforms as the adjoint representation with $V_{\rm R}^{}$, are actually not independent due to the cyclic property of trace. For example, we have ${\rm Tr} \left(M_l^{\dagger}M_l^{}\right)={\rm Tr} \left(M_l M_l^\dagger\right)={\rm Tr}\left(H_l^{}\right)$.}
\begin{eqnarray}
\label{eq:buliding blocks}
H_l^{}&\equiv& M_l^{}M_l^{\dagger}\rightarrow H_l^{\prime}=U_{\rm L}H_l^{}U_{\rm L}^{\dagger}\;,\nonumber\\
H_{\rm D}^{}&\equiv& M_{\rm D}^{}M_{\rm D}^{\dagger}\rightarrow H_{\rm D}^{\prime}=U_{\rm L}^{}H_{\rm D}^{}U_{\rm L}^{\dagger}\;,\nonumber\\
H_{\rm R}^{}&\equiv& M_{\rm R}^{\dagger}M_{\rm R}\rightarrow H_{\rm R}^{\prime}=U_{\rm R}H_{\rm R}^{}U_{\rm R}^{\dagger}\;,\nonumber\\
\tilde{H}_{\rm D}^{}&\equiv& M_{\rm D}^{\dagger}M_{\rm D}^{}\rightarrow \tilde{H}_{\rm D}^{\prime}=U_{\rm R}^{}\tilde{H}_{\rm D}^{}U_{\rm R}^{\dagger}\;,\nonumber\\
G_{\rm DR}^{(n)}&\equiv& M_{\rm R}^{\dagger}\left(\tilde{H}_{\rm D}^{*}\right)_{}^n M_{\rm R}^{}\rightarrow G_{\rm DR}^{(n)\prime}=U_{\rm R}^{}G_{\rm DR}^{}U_{\rm R}^{\dagger}\;,\nonumber\\
G_{l \rm D}^{(n)}&\equiv& M_{\rm D}^{\dagger}\left(H_l^{}\right)_{}^{n}M_{\rm D}^{}\rightarrow G_{l \rm D}^{(n)\prime}=U_{\rm R}^{}G_{l \rm D}^{(n)}U_{\rm R}^{\dagger}\;,\nonumber\\
G_{\rm P}^{}&\equiv& M_{\rm R}^{\dagger}\left\{\,\cdots\right\}M_{\rm R}^{}\rightarrow G_{\rm P}^{\prime}=U_{\rm R}^{}G_{\rm P}^{} U_{\rm R}^{\dagger}\;,
\end{eqnarray}
where $n$ is a positive integer\footnote{It should be noted that for
the building blocks as the adjoint representation of ${\rm U}(N)$, those with $n\geq N$ are no longer independent and can in fact be expressed as a linear combination of the building blocks with $n<N$ using the Cayley-Hamilton theorem. For the same reason, the power index of each building block in Eq.~(\ref{eq:general invariants}) must
be smaller than $N$. These two constraints largely reduce the number of the independent building blocks, rendering it possible to construct all the basic invariants explicitly.} and $G_{\rm P}^{}$ denotes a class of building blocks. To be more specific, the ellipses ``\,$\cdots$" stand for the products of $\tilde{H}_{\rm D}^{*}$ and $G_{l\rm D}^{(n) *}$ to some power. Thus the building blocks in Eq.~(\ref{eq:buliding blocks}) can be divided into two categories, belonging to the adjoint representation of ${\rm U}(3)$ and ${\rm U}(N)$, respectively:
\begin{eqnarray}
A&\rightarrow& U_{\rm L}^{} A U_{\rm L}^{\dagger}\;,\quad
A=\left\{H_l^{},H_{\rm D}^{}\right\}\;,\nonumber\\
B&\rightarrow& U_{\rm R}^{} B U_{\rm R}^{\dagger}\;,\quad
B=\left\{H_{\rm R}^{},\tilde{H}_{\rm D}^{}, G_{\rm DR}^{(n)}, G_{l\rm D}^{(n)},G_{\rm P}^{}\right\}\;.
\end{eqnarray}
As a consequence, one can immediately write down two classes of flavor invariants
\begin{eqnarray}
\label{eq:general invariants}
I_A^{}&=&{\rm Tr}\left(A_i^a A_j^b A_k^c \cdots \right)\;,\quad
A_i^{} ,A_j^{}, A_k^{} \in A\;;
\nonumber\\
I_B^{}&=&{\rm Tr}\left(B_i^d B_j^e B_k^f \cdots \right)\;,\quad
B_i^{} ,B_j^{}, B_k^{} \in B\;,
\end{eqnarray}
where the non-negative integers $\left\{a, b, c, d, e, f\right\}$ stand for the power indices of the corresponding matrices and the ellipses ``\,$\cdots$" denote the additional possible matrices in the set $A$ or $B$. As one can see, the structures of the building blocks in the seesaw model are much more complicated than those in the low-energy effective theory, where all the building blocks transform with the unique unitary matrix $U_{\rm L}^{}$~\cite{Wang2021}. This reflects the fact that richer leptonic flavor structures and a more complicated invariant ring exist in the full seesaw model at the high-energy scale.

\section{Hilbert Series in the Minimal Seesaw Model}
\label{sec:HS}
Given the symmetry group and the representation of the building blocks under the group, it is straightforward to calculate the HS using the MW formula in Eq.~(\ref{eq:Molien-Weyl formula}). From Eq.~(\ref{eq:mass matrices trans}) one can observe that $H_l^{}\equiv M_l^{}M_l^\dagger$ belongs to the adjoint representation of ${\rm U}(3)$, $M_{\rm D}^{}$ to the bi-fundamental representation of ${\rm U}(3)$ and ${\rm U}(N)$, while $M_{\rm R}^{}$ to the rank-two symmetric tensor representation of ${\rm U}(N)$, i.e.,
\begin{eqnarray}
H_l^{}: {\bf 3}_{\rm L}^{}\otimes {\bf 3}_{\rm L}^*\;,~~~~
M_{\rm D}^{}: {\bf 3}_{\rm L}^{} \otimes {\bf N}_{\rm R}^*\;,~~~~
M_{\rm D}^{\dagger}: {\bf N}_{\rm R}^{} \otimes {\bf 3}_{\rm L}^*\;,~~~~
M_{\rm R}^{}: \left({\bf N}_{\rm R}^*\otimes {\bf N}_{\rm R}^*\right)_{\rm s}^{}\;,~~~~ M_{\rm R}^{\dagger}: \left({\bf N}_{\rm R}^{}\otimes {\bf N}_{\rm R}^{}\right)_{\rm s}^{}\;,
\end{eqnarray}
where ${\bf 3}_{\rm L}^{}$ (or ${\bf N}_{\rm R}^{}$) and ${\bf 3}_{\rm L}^{*}$ (or ${\bf N}_{\rm R}^{*}$) denote respectively the fundamental and anti-fundamental representation of ${\rm U}(3)$ (or ${\rm U}(N)$) and the subscript ``s" refers to the symmetric part. Recalling that the character functions of the fundamental and anti-fundamental representation of ${\rm U}(N)$ group are $\sum_{i=1}^{N} z_i^{}$ and $\sum_{i=1}^{N} z_i^{-1}$, respectively, with $z_i^{}$ being the coordinates on the maximum torus of ${\rm U}(N)$~\cite{Wang2021}, one can calculate the character functions of any representations via the tensor product decomposition. In the subsequent two subsections, we will consider two concrete models, namely, the toy model with $N=1$ and the realistic MSM with $N=2$.

\subsection{Toy Model}
As a warm-up exercise, we start with the toy model where there is only one generation of RH neutrino, i.e., $N=1$. In this case, the character functions of $M_{\rm R}^{}$, $H_l^{}$, and $M_{\rm D}^{}$ read
\begin{eqnarray}
\chi_{\rm R}^{}\left(z_4^{}\right) &=& z_4^2 + z_4^{-2} \;, \nonumber \\
\chi_l^{}\left(z_1^{},z_2^{},z_3^{}\right) &=& \left(z_1^{}+z_2^{}+z_3^{}\right)\left(z_1^{-1}+z_2^{-1}+z_3^{-1}\right)\;,
\nonumber\\
\chi_{\rm D}^{}\left(z_1^{},z_2^{},z_3^{},z_4^{}\right) &=& \left(z_1^{}+z_2^{}+z_3^{}\right)z_4^{-1}+\left(z_1^{-1}+z_2^{-1}+z_3^{-1}\right) z_4^{}\;,
\end{eqnarray}
where $z_4^{}$ is the coordinate on the maximum torus of ${\rm U}(1)$ while $z_i^{}$ (for $i=1,2,3$) are those of ${\rm U}(3)$. Labeling the
degrees of $M_l^{}$, $M_{\rm D}^{}$ and $M_{\rm R}^{}$ by $q_l^{}$, $q_{\rm D}^{}$ and $q_{\rm R}$, respectively, one can obtain the PE as below
\begin{eqnarray}
\label{eq:PE toy model}
&{\rm PE}&\left(z_1^{},z_2^{},z_3^{},z_4^{};q_l^{},q_{\rm D}^{},q_{\rm R}^{}\right)={\rm exp}\left(\sum_{k=1}^{\infty}\frac{\chi_l^{}\left(z_1^{k},z_2^{k},z_3^{k}\right) q_l^{2k}+\chi_{\rm D}^{}\left(z_1^{k},z_2^{k},z_3^{k},z_4^{k}\right)q_{\rm D}^{k}+\chi_{\rm R}^{}\left(z_4^{k}\right)q_{\rm R}^k}{k} \right)\nonumber\\
&=&\left[\left(1-q_l^2\right)^3\left(1-q_l^2 z_1^{}z_2^{-1}\right)\left(1-q_l^2 z_2^{}z_1^{-1}\right)\left(1-q_l^2 z_1^{}z_3^{-1}\right)\left(1-q_l^2 z_3^{}z_1^{-1}\right)\left(1-q_l^2 z_2^{}z_3^{-1}\right)\right.\nonumber\\
&& \left. \times \left(1-q_l^2 z_3^{}z_2^{-1}\right)\left(1-q_{\rm D}^{}z_1^{}z_4^{-1}\right)\left(1-q_{\rm D}^{}z_4^{}z_1^{-1}\right)\left(1-q_{\rm D}^{}z_2^{}z_4^{-1}\right)\left(1-q_{\rm D}^{}z_4^{}z_2^{-1}\right)
\left(1-q_{\rm D}^{}z_3^{}z_4^{-1}\right)\right.\nonumber\\
&& \left. \times \left(1-q_{\rm D}^{}z_4^{}z_3^{-1}\right)\left(1-q_{\rm R}^{}z_4^2\right)\left(1-q_{\rm R}^{}z_4^{-2}\right)
\right]_{}^{-1}\;,
\end{eqnarray}
while the Haar measure of ${\rm U}(3)\otimes {\rm U}(1)$ reads
\begin{eqnarray}
\label{eq:Haar measure toy model}
\int \left[{\rm d}\mu\right]_{{\rm U}(3)\otimes {\rm U}(1)}^{}&=&\frac{1}{3!}\left(\frac{1}{2\pi{\rm i}}\right)_{}^4 \oint_{\left|z_1^{}\right|=1}^{}\frac{{\rm d}z_1}{z_1}\oint_{\left|z_2^{}\right|=1}^{}\frac{{\rm d}z_2}{z_2}\oint_{\left|z_3^{}\right|=1}^{}\frac{{\rm d}z_3}{z_3}\nonumber\\
&&\times
\oint_{\left|z_4^{}\right|=1}^{}\frac{{\rm d}z_4}{z_4}\left[-\frac{\left(z_2-z_1\right)^2\left(z_3-z_1\right)^2 \left(z_3-z_2\right)^2}{z_1^2z_2^2z_3^2}
\right]\;.
\end{eqnarray}
Substituting Eqs.~(\ref{eq:PE toy model}) and (\ref{eq:Haar measure toy model}) into the MW formula in Eq.~(\ref{eq:Molien-Weyl formula}) and performing the complex integrals via the residue theorem, one gets the multi-graded HS
\begin{eqnarray}
\label{eq:multi HS toy model}
{\cal H}\left(q_l^{},q_{\rm D}^{},q_{\rm R}^{}\right)&=&\int \left[{\rm d}\mu\right]_{{\rm U}(3)\otimes {\rm U}(1)}^{}{\rm PE}\left(z_1^{},z_2^{},z_3^{},z_4^{};q_l^{},q_{\rm D}^{},q_{\rm R}^{}\right)\nonumber\\
&=&\frac{1}{\left(1-q_l^2\right)\left(1-q_l^4\right)\left(1-q_l^6\right) \left(1-q_{\rm D}^2\right)\left(1-q_l^2q_{\rm D}^2\right)\left(1-q_l^4 q_{\rm D}^2\right)\left(1-q_{\rm R}^2\right)}\;,
\end{eqnarray}
from which we can calculate the PL
\begin{eqnarray}
\label{eq:PL toy model}
{\rm PL}\left[{\cal H}\left(q_l^{},q_{\rm D}^{},q_{\rm R}^{}\right)\right]=q_l^2+q_{\rm D}^2+q_{\rm R}^2+q_l^4+q_l^2q_{\rm D}^2+q_l^6+q_l^4q_{\rm D}^2\;,
\end{eqnarray}
and the ungraded HS
\begin{eqnarray}
\label{eq:ungraded HS toy model}
{\mathscr H}\left(q\right)\equiv{\cal H}\left(q,q,q\right)=\frac{1}{\left(1-q^2\right)^3\left(1-q^4\right)^2 \left(1-q^6\right)^2}\;,
\end{eqnarray}
where the last identity has been derived by identifying $q_l^{} = q_{\rm D}^{}=q_{\rm R}^{}\equiv q$ in Eq.~(\ref{eq:multi HS toy model}). Some comments on the results in Eqs.~(\ref{eq:PL toy model}) and (\ref{eq:ungraded HS toy model}) are in order.
\begin{table}[t!]
\centering
\begin{tabular}{l|c|c}
\hline \hline
flavor invariants &  degree & CP \\
\hline \hline
$J_{200}^{}\equiv {\rm Tr}\left(H_{l}^{}\right)$ &  2 & + \\
\hline
$J_{020}^{}\equiv {\rm Tr}\left(H_{\rm D}^{}\right)$ &  2 & +\\
\hline
$J_{002}^{}\equiv {\rm Tr}\left(H_{\rm R}^{}\right)$ &  2 &+\\
\hline
$J_{400}^{}\equiv {\rm Tr}\left(H_{l}^2\right)$ & 4 &+\\
\hline
$J_{220}^{}\equiv {\rm Tr}\left(H_l^{}H_{\rm D}^{}\right)$ & 4 &+\\
\hline
$J_{600}^{}\equiv {\rm Tr}\left(H_{l}^3\right)$ &  6 &+\\
\hline
$J_{420}^{}\equiv {\rm Tr}\left(H_l^2H_{\rm D}^{}\right)$ &  6 & $+$\\
\hline
\hline
\end{tabular}
\vspace{0.5cm}
\caption{Summary of the basic flavor invariants in the generating set along with their degrees and CP parities in the case of one-generation RH neutrino, where $q_l^{}$, $q_{\rm D}^{}$ and $q_{\rm R}^{}$ denote the degrees of $M_l^{}$, $M_{\rm D}^{}$ and $M_{\rm R}^{}$, respectively.}
\label{table:toy model}
\end{table}
\begin{itemize}
\item The numerator of ${\mathscr H}\left(q\right)$ is simply one, implying that the invariant ring of the toy model is free and there is no syzygy. In other words, all the primary invariants are also the basic invariants in the generating set.
\item From the denominator of ${\mathscr H}\left(q\right)$ we can observe that there are totally 7 primary invariants.  Correspondingly, there are also 7 physical parameters in the model, i.e., three charged-lepton masses, one RH-neutrino mass and the moduli of three elements of the $3\times 1$ Dirac neutrino mass matrix $M^{}_{\rm D}$.\footnote{In the case of one-generation RH neutrino, all the phases in the Yukawa matrix can be absorbed by the SM neutrino fields. As a result, there is no CP violation in the theory, which corresponds to the fact that all the basic invariants in Table~\ref{table:toy model} are CP-even.} Furthermore, the degrees of the primary invariants and the number of primary invariants at a certain degree can also be read off from the denominator of ${\mathscr H}\left(q\right)$ in Eq.~(\ref{eq:ungraded HS toy model}): There are three primary invariants of degree 2, two of degree 4 and two of degree 6.
\item The positive terms in Eq.~(\ref{eq:PL toy model}) show the structures of all the basic invariants and their degrees as well. Then it is easy to construct explicitly all the basic invariants, which are summarized in Table~\ref{table:toy model}. Note that the absence of any negative terms in Eq.~(\ref{eq:PL toy model}) also implies the absence of any syzygies, as expected for a free ring.
\end{itemize}

Thus all the flavor invariants in the toy model with one generation of RH neutrino can be decomposed into the polynomials of the 7 basic invariants in Table~{\ref{table:toy model}}. For example, ${\rm Tr}\left(\tilde{H}_{\rm D}^{}\right)={\rm Tr}\left(H_{\rm D}^{}\right)=J_{020}^{}$ and ${\rm Tr}\left(H_l^{}H_{\rm D}^2\right)={\rm Tr}\left(H_l^{}H_{\rm D}^{}\right) {\rm Tr}\left(\tilde{H}_{\rm D}^{}\right)=J_{220}^{}J_{020}^{}$. A systematic approach to decomposing an arbitrary invariant into the polynomial of the basic invariants and to finding out all the syzygies at a certain degree is presented in Appendix C of Ref.~\cite{Wang2021}.

\subsection{Minimal Seesaw Model}
Now we proceed with the MSM with two generations of RH neutrinos, i.e., $N=2$. In this case, the character functions of $M_{\rm R}^{}$, $H_l^{}$ and $M_{\rm D}^{}$ turn out to be
\begin{eqnarray}
\chi_{\rm R}^{} \left(z_4^{},z_5^{}\right) &=& z_4^2 + z_5^2 + z_4^{}z_5^{} + z_4^{-1}z_5^{-1} + z_4^{-2} + z_5^{-2} \;, \nonumber \\
\chi_l^{}\left(z_1^{},z_2^{},z_3^{}\right) &=& \left(z_1^{}+z_2^{}+z_3^{}\right)\left(z_1^{-1}+z_2^{-1}+z_3^{-1}\right)\;,
\nonumber\\
\chi_{\rm D}^{}\left(z_1^{},z_2^{},z_3^{},z_4^{},z_5^{}\right) &=& \left(z_1^{}+z_2^{}+z_3^{}\right) \left(z_4^{-1}+z_5^{-1}\right) + \left(z_1^{-1}+z_2^{-1}+z_3^{-1}\right) \left(z_4^{}+z_5^{}\right) \;,
\end{eqnarray}
where $z_j^{}$ (for $j=4,5$) are the coordinates on the maximum torus of ${\rm U}(2)$ while $z_i^{}$ (for $i=1,2,3$) are those of ${\rm U}(3)$. Then the PE becomes
\begin{eqnarray}
\label{eq:PE MSM}
&{\rm PE}&\left(z_1^{},z_2^{},z_3^{},z_4^{},z_5^{};q_l^{},q_{\rm D}^{},q_{\rm R}^{}\right)\nonumber\\
&=&{\rm exp}\left(\sum_{k=1}^{\infty}\frac{\chi_l^{}\left(z_1^{k},z_2^{k},z_3^{k}\right) q_l^{2k}+\chi_{\rm D}^{}\left(z_1^{k},z_2^{k},z_3^{k},z_4^{k},z_5^{k}\right)q_{\rm D}^{k}+\chi_{\rm R}^{}\left(z_4^{k},z_5^{k}\right)q_{\rm R}^k}{k} \right)\nonumber\\
&=&\left[\left(1-q_l^2\right)^3\left(1-q_l^2 z_1^{}z_2^{-1}\right)\left(1-q_l^2 z_2^{}z_1^{-1}\right)\left(1-q_l^2 z_1^{}z_3^{-1}\right)\left(1-q_l^2 z_3^{}z_1^{-1}\right)\left(1-q_l^2 z_2^{}z_3^{-1}\right)\right.\nonumber\\
&& \left. \times \left(1-q_l^2 z_3^{}z_2^{-1}\right)\left(1-q_{\rm D}^{}z_1^{}z_4^{-1}\right)\left(1-q_{\rm D}^{}z_4^{}z_1^{-1}\right)\left(1-q_{\rm D}^{}z_2^{}z_4^{-1}\right)\left(1-q_{\rm D}^{}z_4^{}z_2^{-1}\right)
\left(1-q_{\rm D}^{}z_3^{}z_4^{-1}\right)\right.\nonumber\\
&& \left. \times \left(1-q_{\rm D}^{}z_4^{}z_3^{-1}\right)
\left(1-q_{\rm D}^{}z_1^{}z_5^{-1}\right)\left(1-q_{\rm D}^{}z_5^{}z_1^{-1}\right)\left(1-q_{\rm D}^{}z_2^{}z_5^{-1}\right)\left(1-q_{\rm D}^{}z_5^{}z_2^{-1}\right)
\left(1-q_{\rm D}^{}z_3^{}z_5^{-1}\right)\right.\nonumber\\
&&\left. \times \left(1-q_{\rm D}^{}z_5^{}z_3^{-1}\right)\left(1-q_{\rm R}^{}z_4^2\right)\left(1-q_{\rm R}^{} z_4^{}z_5^{}\right)\left(1-q_{\rm R}^{}z_5^2\right)\left(1-q_{\rm R}^{}z_4^{-2}\right)\left(1-q_{\rm R}^{} z_4^{-1}z_5^{-1}\right)\right.\nonumber\\
&&\left.\times \left(1-q_{\rm R}^{}z_5^{-2}\right)
\right]_{}^{-1}\;,
\end{eqnarray}
where the degrees of $M_l^{}$, $M_{\rm D}^{}$ and $M_{\rm R}^{}$ are labeled by $q_l^{}$, $q_{\rm D}^{}$ and $q_{\rm R}$, respectively. The Haar measure of the direct product of two groups ${\rm U}(3)\otimes {\rm U}(2)$
can be written as
\begin{eqnarray}
\label{eq:Haar measure MSM}
\int \left[{\rm d}\mu\right]_{{\rm U}(3)\otimes {\rm U}(2)}^{}&=&\frac{1}{12}\left(\frac{1}{2\pi{\rm i}}\right)_{}^5 \oint_{\left|z_1^{}\right|=1}^{}\frac{{\rm d}z_1}{z_1}\oint_{\left|z_2^{}\right|=1}^{}\frac{{\rm d}z_2}{z_2}\oint_{\left|z_3^{}\right|=1}^{}\frac{{\rm d}z_3}{z_3}\oint_{\left|z_4^{}\right|=1}^{}\frac{{\rm d}z_4}{z_4}\nonumber\\
&&\times \oint_{\left|z_5^{}\right|=1}^{}\frac{{\rm d}z_5}{z_5}\left[-\frac{\left(z_2-z_1\right)^2\left(z_3-z_1\right)^2 \left(z_3-z_2\right)^2}{z_1^2z_2^2z_3^2}
\right]\left(2-\frac{z_4}{z_5}-\frac{z_5}{z_4}\right)\;.
\end{eqnarray}

Inserting Eqs.~(\ref{eq:PE MSM}) and (\ref{eq:Haar measure MSM}) into the MW formula in Eq.~(\ref{eq:Molien-Weyl formula}) and calculating the complex integrals by virtue of the residue theorem, we get the multi-graded HS in the MSM
\begin{eqnarray}
\label{eq:multi HS MSM}
{\cal H}\left(q_l^{},q_{\rm D}^{},q_{\rm R}^{}\right)=\int \left[{\rm d}\mu\right]_{{\rm U}(3)\otimes {\rm U}(2)}^{}{\rm PE}\left(z_1^{},z_2^{},z_3^{},z_4^{},z_5^{};q_l^{},q_{\rm D}^{},q_{\rm R}^{}\right)=\frac{{\cal N}\left(q_l^{},q_{\rm D}^{},q_{\rm R}^{}\right)}{{\cal D}\left(q_l^{},q_{\rm D}^{},q_{\rm R}^{}\right)}\;,
\end{eqnarray}
where
{\allowdisplaybreaks
\begin{eqnarray*}
{\cal N}\left(q_l^{},q_{\rm D}^{},q_{\rm R}^{}\right)&=&
-q_{\rm D}^{22} q_{\rm R}^{12} q_l^{22} - q_{\rm D}^{18} q_{\rm R}^8 q_l^{22} + q_{\rm D}^{20} q_{\rm R}^{12} q_l^{20} - 2 q_{\rm D}^{18}q_{\rm R}^{10} q_l^{20} - q_{\rm D}^{16} q_{\rm R}^{10} q_l^{20} - q_{\rm D}^{18} q_{\rm R}^8 q_l^{20} + q_{\rm D}^{16} q_{\rm R}^8 q_l^{20}\\
&& + q_{\rm D}^{14} q_{\rm R}^8 q_l^{20} + q_{\rm D}^{16} q_{\rm R}^6 q_l^{20} - q_{\rm D}^{18} q_{\rm R}^{12} q_l^{18} - 2
q_{\rm D}^{18} q_{\rm R}^{10} q_l^{18} + q_{\rm D}^{14} q_{\rm R}^{10} q_l^{18} - 2 q_{\rm D}^{18} q_{\rm R}^8 q_l^{18} + q_{\rm D}^{16}
q_{\rm R}^8 q_l^{18}\\
&& + q_{\rm D}^{14} q_{\rm R}^8 q_l^{18} - q_{\rm D}^{12} q_{\rm R}^8 q_l^{18} + 2 q_{\rm D}^{16} q_{\rm R}^6 q_l^{18} + q_{\rm D}^{14} q_{\rm R}^6 q_l^{18} - 2 q_{\rm D}^{18} q_{\rm R}^{10} q_l^{16} + q_{\rm D}^{14} q_{\rm R}^{10} q_l^{16} - q_{\rm D}^{18} q_{\rm R}^8 q_l^{16}\\
&& + 3 q_{\rm D}^{16} q_{\rm R}^8 q_l^{16} + 3 q_{\rm D}^{14} q_{\rm R}^8
q_l^{16} - q_{\rm D}^{12} q_{\rm R}^8 q_l^{16} + 2 q_{\rm D}^{16} q_{\rm R}^6 q_l^{16} - q_{\rm D}^{14} q_{\rm R}^6 q_l^{16} - 2 q_{\rm D}^{12} q_{\rm R}^6 q_l^{16} - q_{\rm D}^{14} q_{\rm R}^4 q_l^{16}\\
&& + q_{\rm D}^{10} q_{\rm R}^4 q_l^{16} + q_{\rm D}^{14} q_{\rm R}^{10}
   q_l^{14} - q_{\rm D}^{18} q_{\rm R}^8 q_l^{14} + q_{\rm D}^{16} q_{\rm R}^8 q_l^{14} + q_{\rm D}^{14} q_{\rm R}^8 q_l^{14} - 2 q_{\rm D}^{12} q_{\rm R}^8 q_l^{14} + 2 q_{\rm D}^{16} q_{\rm R}^6 q_l^{14}\\
&& + q_{\rm D}^{14} q_{\rm R}^6 q_l^{14} + 2 q_{\rm D}^{10} q_{\rm R}^6
q_l^{14} - q_{\rm D}^{14} q_{\rm R}^4 q_l^{14} + q_{\rm D}^{10} q_{\rm R}^4 q_l^{14} - q_{\rm D}^8 q_{\rm R}^4 q_l^{14} - q_{\rm D}^{16}q_{\rm R}^{10} q_l^{12} + q_{\rm D}^{14} q_{\rm R}^{10} q_l^{12}\\
&& + q_{\rm D}^{12} q_{\rm R}^{10} q_l^{12} + q_{\rm D}^{16} q_{\rm R}^8
q_l^{12} + 3 q_{\rm D}^{14} q_{\rm R}^8 q_l^{12} + q_{\rm D}^{16} q_{\rm R}^6 q_l^{12} - q_{\rm D}^{14} q_{\rm R}^6 q_l^{12} - 3
q_{\rm D}^{12} q_{\rm R}^6 q_l^{12} - q_{\rm D}^8 q_{\rm R}^6 q_l^{12}\\
&& - q_{\rm D}^{14} q_{\rm R}^4 q_l^{12} + 2 q_{\rm D}^{10} q_{\rm R}^4
q_l^{12} - q_{\rm D}^8 q_{\rm R}^4 q_l^{12}- q_{\rm D}^8 q_{\rm R}^2 q_l^{12} + q_{\rm D}^{14} q_{\rm R}^{10} q_l^{10} + q_{\rm D}^{14}
q_{\rm R}^8 q_l^{10} - 2 q_{\rm D}^{12} q_{\rm R}^8 q_l^{10}\\
&& + q_{\rm D}^8 q_{\rm R}^8 q_l^{10} + q_{\rm D}^{14} q_{\rm R}^6 q_l^{10} + 3 q_{\rm D}^{10} q_{\rm R}^6 q_l^{10} + q_{\rm D}^8 q_{\rm R}^6 q_l^{10} - q_{\rm D}^6 q_{\rm R}^6 q_l^{10} - 3 q_{\rm D}^8 q_{\rm R}^4
q_l^{10} - q_{\rm D}^6 q_{\rm R}^4 q_l^{10}\\
&& - q_{\rm D}^{10} q_{\rm R}^2 q_l^{10}-q_{\rm D}^8 q_{\rm R}^2 q_l^{10} + q_{\rm D}^6 q_{\rm R}^2 q_l^{10} + q_{\rm D}^{14} q_{\rm R}^8 q_l^8-q_{\rm D}^{12} q_{\rm R}^8 q_l^8 + q_{\rm D}^8 q_{\rm R}^8 q_l^8-2 q_{\rm D}^{12} q_{\rm R}^6 q_l^8\\
&& -q_{\rm D}^8 q_{\rm R}^6 q_l^8 - 2 q_{\rm D}^6 q_{\rm R}^6 q_l^8 + 2 q_{\rm D}^{10} q_{\rm R}^4 q_l^8 - q_{\rm D}^8 q_{\rm R}^4 q_l^8 - q_{\rm D}^6 q_{\rm R}^4 q_l^8 + q_{\rm D}^4 q_{\rm R}^4 q_l^8 - q_{\rm D}^8 q_{\rm R}^2 q_l^8\\
&& - q_{\rm D}^{12} q_{\rm R}^8 q_l^6 + q_{\rm D}^8 q_{\rm R}^8 q_l^6 + 2
q_{\rm D}^{10} q_{\rm R}^6 q_l^6 + q_{\rm D}^8 q_{\rm R}^6 q_l^6 - 2 q_{\rm D}^6 q_{\rm R}^6 q_l^6 + q_{\rm D}^{10} q_{\rm R}^4 q_l^6-3 q_{\rm D}^8 q_{\rm R}^4 q_l^6 \\
&& - 3 q_{\rm D}^6 q_{\rm R}^4 q_l^6 + q_{\rm D}^4 q_{\rm R}^4 q_l^6-q_{\rm D}^8 q_{\rm R}^2 q_l^6 + 2 q_{\rm D}^4 q_{\rm R}^2 q_l^6-q_{\rm D}^8 q_{\rm R}^6 q_l^4 - 2 q_{\rm D}^6 q_{\rm R}^6 q_l^4 + q_{\rm D}^4 q_l^4\\
&& + q_{\rm D}^{10} q_{\rm R}^4 q_l^4 - q_{\rm D}^8 q_{\rm R}^4
   q_l^4 - q_{\rm D}^6 q_{\rm R}^4 q_l^4 + 2 q_{\rm D}^4 q_{\rm R}^4 q_l^4 - q_{\rm D}^8 q_{\rm R}^2 q_l^4 + 2 q_{\rm D}^4 q_{\rm R}^2 q_l^4 - q_{\rm D}^6 q_{\rm R}^6 q_l^2 \\
&& - q_{\rm D}^8 q_{\rm R}^4 q_l^2 - q_{\rm D}^6 q_{\rm R}^4 q_l^2 + q_{\rm D}^4 q_{\rm R}^4 q_l^2 - q_{\rm D}^2 q_l^2 + q_{\rm D}^6 q_{\rm R}^2 q_l^2 + 2 q_{\rm D}^4 q_{\rm R}^2 q_l^2 + q_{\rm D}^4 q_{\rm R}^4 + 1
\;,\\
{\cal D}\left(q_l^{},q_{\rm D}^{},q_{\rm R}^{}\right)&=&\left(1-q_l^2\right)\left(1-q_{\rm D}^2\right)\left(1-q_{\rm R}^2\right)\left(1-q_l^4\right)\left(1-q_{\rm D}^4\right)\left(1-q_{\rm R}^4\right)\left(1-q_l^2q_{\rm D}^2\right)_{}^2\left(1-q_{\rm D}^2q_{\rm R}^2\right)\nonumber\\
&&\times \left(1-q_l^6\right)\left(1-q_l^4q_{\rm D}^2\right)\left(1-q_l^2 q_{\rm D}^4\right)\left(1-q_l^2q_{\rm D}^2q_{\rm R}^2\right)\left(1-q_{\rm D}^4 q_{\rm R}^2\right)\left(1-q_l^4q_{\rm D}^2q_{\rm R}^2\right)\nonumber\\
&&\times \left(1-q_l^4 q_{\rm D}^4 q_{\rm R}^2\right)\left(1-q_l^8q_{\rm D}^4q_{\rm R}^2\right)
\;.
\end{eqnarray*}}
From the multi-grade HS in Eq.~(\ref{eq:multi HS MSM}) we can compute the PL
\begin{eqnarray}
\label{eq:PL MSM}
{\rm PL}\left[{\cal H}\left(q_l^{},q_{\rm D}^{},q_{\rm R}^{}\right)\right]&=&\left(q_l^2+q_{\rm D}^2+q_{\rm R}^2\right)+\left(q_l^4+q_l^2 q_{\rm D}^2+q_{\rm D}^4+q_{\rm D}^2q_{\rm R}^2+q_{\rm R}^4\right)+\left(q_l^6+q_l^4 q_{\rm D}^2+q_l^2 q_{\rm D}^4\right. \nonumber\\
&& \left. +q_l^2q_{\rm D}^2q_{\rm R}^2+q_{\rm D}^4q_{\rm R}^2\right)+\left(q_l^4 q_{\rm D}^4+q_l^4 q_{\rm D}^2 q_{\rm R}^2+2q_l^2 q_{\rm D}^4 q_{\rm R}^2+q_{\rm D}^4 q_{\rm R}^4\right)+\left(3q_l^4q_{\rm D}^4 q_{\rm R}^2\right.\nonumber\\
&&\left. +q_l^2 q_{\rm D}^6q_{\rm R}^2+q_l^2q_{\rm D}^4 q_{\rm R}^4\right)+\left(q_l^6q_{\rm D}^6+2q_l^6q_{\rm D}^4q_{\rm R}^2+2q_l^4 q_{\rm D}^6q_{\rm R}^2+2q_l^4 q_{\rm D}^4 q_{\rm R}^4\right)+\left(
q_l^8 q_{\rm D}^4 q_{\rm R}^2\right.\nonumber\\
&& \left. +2q_l^6 q_{\rm D}^6 q_{\rm R}^2+q_l^6 q_{\rm D}^4 q_{\rm R}^4
-q_l^2 q_{\rm D}^8q_{\rm R}^4-q_l^2 q_{\rm D}^6q_{\rm R}^6\right)+\left(2q_l^8 q_{\rm D}^6q_{\rm R}^2+q_l^8 q_{\rm D}^4q_{\rm R}^4-q_l^6 q_{\rm D}^8q_{\rm R}^2\right.\nonumber\\
&&\left. -5q_l^4 q_{\rm D}^8q_{\rm R}^4-q_l^6 q_{\rm D}^6q_{\rm R}^4-2 q_l^2 q_{\rm D}^8q_{\rm R}^6  -2q_l^4 q_{\rm D}^6q_{\rm R}^6- q_{\rm D}^8q_{\rm R}^8\right)+\left(q_l^{10} q_{\rm D}^6 q_{\rm R}^2-q_l^6 q_{\rm D}^{10}q_{\rm R}^2\right.\nonumber\\
&& \left. -q_l^8 q_{\rm D}^8q_{\rm R}^2-2q_l^4 q_{\rm D}^{10}q_{\rm R}^4-8q_l^6 q_{\rm D}^8q_{\rm R}^4-q_l^2 q_{\rm D}^{10} q_{\rm R}^6-6q_l^4 q_{\rm D}^8q_{\rm R}^6-2q_l^6 q_{\rm D}^6q_{\rm R}^6-q_l^2 q_{\rm D}^8q_{\rm R}^8\right)\nonumber\\
&& -{\cal O}\left(\left[q_l^{}q_{\rm D} q_{\rm R}^{}\right]_{}^{20}\right)
\;,
\end{eqnarray}
and the ungraded HS
\begin{eqnarray}
\label{eq:ungraded HS MSM}
{\mathscr H}\left(q\right)\equiv{\cal H}\left(q,q,q\right)=\frac{{\mathscr N}\left(q\right)}{{\mathscr D}\left(q\right)}\;,
\end{eqnarray}
with
\begin{eqnarray*}
{\mathscr N}\left(q\right) &=& 1+q_{}^2+q_{}^4+2q_{}^6+6q_{}^8+10q_{}^{10}+18q_{}^{12}+23q_{}^{14}+28q_{}^{16} +31q_{}^{18}+34q_{}^{20}+32q_{}^{22} \nonumber\\
&&+34q_{}^{24}+31q_{}^{26}+28q_{}^{28}+23q_{}^{30}+18q_{}^{32}+10q_{}^{34} +6q_{}^{36}+2q_{}^{38}+q_{}^{40}+q_{}^{42}+q_{}^{44} \; ,
\end{eqnarray*}
and
\begin{eqnarray*}
{\mathscr D}\left(q\right) &=& \left(1-q_{}^2\right)_{}^2 \left(1-q_{}^4\right)_{}^5 \left(1-q_{}^6\right)_{}^4 \left(1-q_{}^8\right) \left(1-q_{}^{10}\right) \left(1-q_{}^{14}\right)\;.
\end{eqnarray*}
\renewcommand\arraystretch{1.6}
\begin{table}[H]
\centering
\begin{tabular}{l|c|c||l|c|c}
\hline \hline
flavor invariants & degree & CP &flavor invariants & degree & CP \\
\hline \hline
$I_{200}^{}\equiv {\rm Tr}\left(H_{l}^{}\right)$ & 2 & $+$&
$I_{442}^{(2)}\equiv {\rm Tr}\left(G_{\rm DR}^{}G_{l\rm D}^{(2)}\right)$ & 10 & $+$ \\
\hline
$I_{020}^{}\equiv {\rm Tr}\left(H_{\rm D}^{}\right)$ & 2 & $+$ &
$I_{442}^{(3)}\equiv {\rm Tr}\left(\left[\tilde{H}_{\rm D}^{},H_{\rm R}^{} \right]G_{l\rm D}^{(2)}\right)$ & 10 & $-$\\
\hline
$I_{002}^{}\equiv {\rm Tr}\left(H_{\rm R}^{}\right)$ & 2 & $+$ &
$I_{262}^{}\equiv {\rm Tr}\left(\left[\tilde{H}_{\rm D}^{},G_{l\rm D}^{} \right]G_{\rm DR}^{}\right)$ & 10 & $-$ \\
\hline
$I_{400}^{}\equiv {\rm Tr}\left(H_{l}^{2}\right)$&4 & $+$ & $I_{244}^{}\equiv {\rm Tr}\left(\left[H_{\rm R}^{},G_{l\rm D}^{}\right]G_{\rm DR}^{}\right)$ & 10 & $-$ \\
\hline
$I_{220}^{}\equiv {\rm Tr}\left(H_{l}^{}H_{\rm D}^{}\right)$ & 4 & $+$&
$I_{660}^{}\equiv {\rm Tr}\left(\left[\tilde{H}_{\rm D}^{},G_{l\rm D}^{}\right]G_{l\rm D}^{(2)}\right)$ & 12 & $-$ \\
\hline
$I_{040}^{}\equiv {\rm Tr}\left(H_{\rm D}^{2}\right)$ & 4 & $+$ & $I_{642}^{(1)}\equiv {\rm Tr}\left(\left[H_{\rm R}^{},G_{l\rm D}^{}\right]G_{l\rm D}^{(2)}\right)$ & 12 & $-$ \\
\hline
$I_{022}^{}\equiv {\rm Tr}\left(\tilde{H}_{\rm D}^{}H_{\rm R}^{}\right)$ & 4 & $+$ & $I_{642}^{(2)}\equiv{\rm Tr}\left(G_{l\rm D}^{}G_{l\rm DR}^{(2)}\right)$ & 12 & $+$ \\
\hline
$I_{004}^{}\equiv{\rm Tr}\left(H_{\rm R}^2\right)$ & 4 & $+$ & $I_{462}^{(1)}\equiv{\rm Tr}\left(\left[\tilde{H}_{\rm D}^{},G_{l\rm D}^{}\right]G_{l\rm DR}^{}\right)$  & 12& $-$ \\
\hline
$I_{600}^{}\equiv {\rm Tr}\left(H_{l}^{3}\right)$  & 6 & $+$ &
$I_{462}^{(2)}\equiv{\rm Tr}\left(\left[\tilde{H}_{\rm D}^{},G_{\rm DR}^{}\right]G_{l\rm D}^{(2)}\right)$ & 12 & $-$ \\
\hline
$I_{420}^{}\equiv {\rm Tr}\left(H_{l}^2 H_{\rm D}^{} \right)$ & 6 & $+$ &
$I_{444}^{(1)}\equiv {\rm Tr}\left(\left[H_{\rm R}^{},G_{l\rm D}^{}\right]G_{l\rm DR}^{}\right)$  & 12 & $-$ \\
\hline
$I_{240}^{}\equiv{\rm Tr}\left(H_{l}^{} H_{\rm D}^2 \right)$ & 6 & $+$&
$I_{444}^{(2)}\equiv {\rm Tr}\left(\left[H_{\rm R}^{},G_{\rm DR}^{}\right]G_{l\rm D}^{(2)}\right)$  & 12 & $-$  \\
\hline
$I_{222}^{}\equiv {\rm Tr}\left(H_{\rm R}^{} G_{l\rm D}^{}\right)$  & 6 & $+$ & $I_{842}^{}\equiv{\rm Tr}\left(G_{l\rm D}^{(2)}G_{l\rm DR}^{(2)}\right)$ & 14 & $+$\\
\hline
$I_{042}^{}\equiv {\rm Tr}\left(\tilde{H}_{\rm D}^{} G_{\rm DR}^{}\right)$ & 6 & $+$& $I_{662}^{(1)}\equiv {\rm Tr}\left(\left[\tilde{H}_{\rm D}^{},G_{l\rm D}^{(2)}\right]G_{l\rm DR}^{}\right)$ & 14 & $-$  \\
\hline
$I_{440}^{}\equiv {\rm Tr}\left(H_{l}^{2}H_{\rm D}^{2}\right)$  & 8 & $+$ &$I_{662}^{(2)}\equiv{\rm Tr}\left(\left[\tilde{H}_{\rm D}^{},G_{l\rm D}^{}\right]G_{l\rm DR}^{(2)}\right)$  & 14 & $-$  \\
\hline
$I_{422}^{}\equiv{\rm Tr}\left(H_{\rm R}^{}G_{l\rm D}^{(2)}\right)$ & 8 & $+$ & $I_{644}^{}\equiv{\rm Tr}\left(\left[H_{\rm R}^{},G_{l\rm D}^{(2)}\right]G_{l\rm DR}^{}\right)$ & 14 & $-$ \\
\hline
$I_{242}^{(1)}\equiv {\rm Tr}\left(G_{l\rm D}^{}G_{\rm DR}^{}\right)$ & 8 & $+$ &$I_{862}^{(1)}\equiv {\rm Tr}\left(\left[\tilde{H}_{\rm D}^{},G_{l\rm D}^{(2)}\right]G_{l\rm DR}^{(2)}\right) $ & 16 & $-$\\
\hline
$I_{242}^{(2)}\equiv{\rm Tr}\left(\left[H_{\rm R}^{},\tilde{H}_{\rm D}^{}\right]G_{l\rm D}^{}\right)$ & 8 & $-$&
$I_{862}^{(2)}\equiv {\rm Tr}\left(\left[G_{l\rm D}^{},G_{l\rm D}^{(2)}\right]G_{l\rm DR}^{}\right)$ & 16 & $-$ \\
\hline
$I_{044}^{}\equiv {\rm Tr}\left(\left[H_{\rm R}^{},\tilde{H}_{\rm D}^{}\right]G_{\rm DR}^{}\right)$  & 8 & $-$&
$I_{844}^{}\equiv {\rm Tr}\left(\left[H_{\rm R}^{},G_{l\rm D}^{(2)}\right]G_{l\rm DR}^{(2)}\right) $ & 16 & $-$ \\
\hline
$I_{442}^{(1)}\equiv {\rm Tr}\left(G_{l\rm D}^{}G_{l\rm DR}^{}\right)$ & 10 & $+$&$I_{10,6,2}^{}\equiv {\rm Tr}\left(\left[G_{l\rm D}^{},G_{l\rm D}^{(2)}\right]G_{l\rm DR}^{(2)}\right)$ & 18 & $-$ \\
\hline
\hline
\end{tabular}
\vspace{0.5cm}
\caption{
Summary of the basic flavor invariants in the generating set along with their degrees and CP parities in the MSM with two-generation RH neutrinos, where $q_l^{}$, $q_{\rm D}^{}$ and $q_{\rm R}^{}$ denote the degrees of $M_l^{}$, $M_{\rm D}^{}$ and $M_{\rm R}^{}$, respectively. Note that the commutator $\left[A,B\right]\equiv AB-BA$ of two matrices has been defined. To simplify the notations, we have also defined $G_{l\rm D}^{}\equiv G_{l\rm D}^{(1)}$, $G_{\rm DR}^{}\equiv G_{\rm DR}^{(1)}$, $G_{l\rm DR}^{}\equiv M_{\rm R}^{\dagger}G_{l\rm D}^{*}M_{\rm R}^{}$ and $G_{l\rm DR}^{(2)}\equiv M_{\rm R}^{\dagger}\left(G_{l\rm D}^{(2)}\right)_{}^{*}M_{\rm R}^{}$. Among all the 38 basic invariants, 18 are CP-odd and the others are CP-even.}
\label{table:MSM}
\end{table}
\renewcommand\arraystretch{1}

From the results of Eqs.~(\ref{eq:PL MSM}) and (\ref{eq:ungraded HS MSM}) we can extract very important information about the primary and basic invariants. Some helpful comments are in order.

First, from the denominator of ${\mathscr H}\left(q\right)$ in Eq.~(\ref{eq:ungraded HS MSM}), we conclude that there are in total 14 primary invariants, where two of them are of degree 2, five of degree 4, four of degree 6, one of degree 8, one of degree 10 and one of degree 14. On the other hand, without the loss of generality one can always choose the flavor basis where the mass matrices of charged leptons and RH neutrinos are real and diagonal, and then the Dirac neutrino mass matrix can be parametrized as~\cite{Casas2001}
\begin{eqnarray}
\label{eq:CI parametrization}
M_{\rm D}^{}={\rm i} V \sqrt{\widehat{M}_{\nu}^{}}R\sqrt{\widehat{M}_{\rm R}^{}}\;,
\end{eqnarray}
where both the light and heavy Majorana neutrino mass matrices $\widehat{M}_{\nu}^{}={\rm Diag}\left\{0,m_2^{},m_3^{}\right\}$ and $\widehat{M}_{\rm R}^{}={\rm Diag}\left\{M_1^{},M_2^{}\right\}$ are real and diagonal.\footnote{Here we assume neutrino mass ordering to be normal, so the lightest neutrino mass $m_1^{}=0$ is vanishing in the MSM. Accordingly, there is only one Majorana CP-violating phase in the PMNS matrix, i.e., the relative phase between two massive neutrino states. The case of inverted neutrino mass ordering can be similarly analyzed.} The Pontecorvo-Maki-Nakagawa-Sakata (PMNS) matrix~\cite{Pontecorvo1957,MNS1962} $V$ can be decomposed as $V=V_{}^{\prime}\cdot {\rm Diag}\left\{1, {\rm e}_{}^{{\rm i}\sigma}, 1\right\}$, where $V_{}^{\prime}$ is a Cabibbo-Kobayashi-Maskawa (CKM)-like matrix that contains one Dirac-type CP phase $\delta$ and three flavor mixing angles $\left\{\theta_{12}^{},\theta_{13}^{},\theta_{23}^{}\right\}$. In addition, the complex and orthogonal matrix $R$, fulfilling the conditions $R_{}^{\rm T}R={\rm Diag}\left\{1,1\right\}$ and $RR_{}^{\rm T}={\rm Diag}\left\{0,1,1\right\}$, can be written as~\cite{Casas2001}
\begin{eqnarray}
R=\left(
\begin{matrix}
0&0\\
\cos z& -\sin z\\
\pm\sin z & \pm\cos z
\end{matrix}
\right)\;,
\end{eqnarray}
with $z$ being an arbitrary complex number. From this parametrization it is clear that there are 14 physical observables in the theory, i.e.,
three charged-lepton masses $\{m^{}_e, m^{}_\mu, m^{}_\tau\}$, two RH neutrino masses $\{M_1^{},M_2^{}\}$, two light neutrino masses $\{m_2^{},m_3^{}\}$, three mixing angles $\left\{\theta_{12}^{},\theta_{13}^{},\theta_{23}^{}\right\}$, one Dirac-type CP phase $\delta$, one Majorana-type CP phase $\sigma$ and the real and imaginary parts of one complex parameter $\{{\rm Re}z,{\rm Im}z\}$. Therefore, we have verified that the number of the primary invariants in the ring is equal to the number of the physical parameters in the theory.

Second, from the first positive terms of PL in Eq.~(\ref{eq:PL MSM}), there are totally 38 basic invariants in the generating set and we have explicitly constructed them, which together with their degrees and CP parities are summarized in Table~\ref{table:MSM}. Among them, 20 invariants are CP-even and the others are CP-odd. On this point, it is worth emphasizing that for the multi-graded PL in the case of the non-complete intersection ring, the ``leading positive terms" should refer to all the positive terms before the first \emph{purely} negative total degree.\footnote{For example, in Eq.~(\ref{eq:PL MSM}), the first purely negative total degree is 20. The reason is that for any total degree lower than 20 there are positive terms at the same degree, whereas all the terms with total degree of 20 are negative.} This observation has not been made explicitly in the literature, to the best of our knowledge, although it has been verified through several concrete examples in Ref.~ \cite{Hanany2008}.

Third, with the renormalization-group equations (RGEs) of $M_l^{}$, $M_{\rm D}^{}$ and $M_{\rm R}^{}$ in the seesaw model~\cite{Haba1998,Casas1999}, it is straightforward to calculate the RGEs of all the flavor invariants. In fact, we have derived the RGEs of the 38 basic invariants listed in Table~\ref{table:MSM} and verified that they form a closed system of differential equations. Such calculations strengthen our belief in the completeness of the generating set. Furthermore, we have also checked the independence of all the basic invariants using the method developed in Appendix C of Ref.~\cite{Wang2021}. We find that none of the basic invariants can be written as the polynomial of the other 37 invariants, indicating that there is no redundancy in the basic invariants in the generating set.

\section{Further Discussions}
\label{sec:discussion}
With all the basic invariants in the MSM, we explore their relations to the flavor invariants in the low-energy effective theory in this section. Furthermore, we also discuss some phenomenological applications of the flavor invariants in the MSM, such as the sufficient and necessary conditions for CP conservation and the CP asymmetries in the decays of heavy Majorana neutrinos.

\subsection{Connection between Low- and High-scale Invariants}
\label{subsec:relation}
In Ref.~\cite{Wang2021}, we obtain all the basic flavor invariants $\left\{I_1^{},I_2^{},...,I_{34}^{}\right\}$ in the low-energy effective theory of three generations of light Majorana neutrinos. Here the notations of flavor invariants in Ref.~\cite{Wang2021} in the low-energy effective theory are followed. An intriguing question is how the basic flavor invariants in the full seesaw model are related to those in the low-energy effective theory.

In the MSM, all the flavor invariants are built upon the charged-lepton mass matrix $M_l^{}$, the Dirac neutrino mass matrix $M_{\rm D}^{}$ and the RH neutrino mass matrix $M_{\rm R}^{}$. As the seesaw scale is usually much higher than the electroweak scale, one can integrate out RH neutrinos and thus obtain the effective neutrino mass matrix in Eq.~(\ref{eq:seesaw formula}), which, together with $M_l^{}$, serves as the building blocks of the flavor invariants in the low-energy effective theory. Interestingly, we find that all the flavor invariants at the low-energy scale can be expressed as the \emph{rational} functions of those at the high-energy scale. This can be realized by simply replacing $M_\nu^{}$ with the right-hand side of Eq.~(\ref{eq:seesaw formula}) and utilizing the following identity
\begin{eqnarray}
\label{eq:matching identity}
A_{}^{-1}=\frac{2\left[{\rm Tr}\left(A\right){\bf 1}_2-A\right]}{{\rm Tr}\left(A\right)^2-{\rm Tr}\left(A^2\right)}\;,
\end{eqnarray}
with $A$ being an arbitrary $2\times2$ non-singular matrix and ${\bf 1}_2^{}$ being the two-dimensional identity matrix. For instance, $I_2^{}\equiv {\rm Tr}\left(H_\nu^{}\right)\equiv {\rm Tr}\left(M_\nu^{} M_\nu^{\dagger}\right)$ can be rewritten as
\begin{eqnarray}
\label{eq:match Hnu}
I_2^{}&=&{\rm Tr}\left[M_{\rm D}^{}M_{\rm R}^{-1}M_{\rm D}^{\rm T}M_{\rm D}^{*}\left(M_{\rm R}^{\dagger}\right)_{}^{-1}M_{\rm D}^{\dagger}\right]={\rm Tr}\left\{\tilde{H}_{\rm D}^{}\left[M_{\rm R}^{\dagger}\left(\tilde{H}_{\rm D}^{*}\right)_{}^{-1} M_{\rm R}\right]_{}^{-1}\right\}\nonumber\\
&=&2\left(I_{002}^{}I_{020}^2-2I_{020}^{}I_{022}^{}+I_{042}^{}\right) /\left(I_{002}^2-I_{004}\right)\;.
\end{eqnarray}
As for $I_5^{}\equiv {\rm Tr}\left(H_\nu^2\right)$, we have
\begin{eqnarray}
\label{eq:match Hnu2}
I_5^{}&=&{\rm Tr}\left[M_{\rm D}^{}M_{\rm R}^{-1}M_{\rm D}^{\rm T}M_{\rm D}^{*}\left(M_{\rm R}^{\dagger}\right)_{}^{-1}M_{\rm D}^{\dagger}M_{\rm D}^{}M_{\rm R}^{-1}M_{\rm D}^{\rm T}M_{\rm D}^{*}\left(M_{\rm R}^{\dagger}\right)_{}^{-1}M_{\rm D}^{\dagger}\right]\nonumber\\
&=&{\rm Tr}\left\{\tilde{H}_{\rm D}^{}\left[M_{\rm R}^{\dagger}\left(\tilde{H}_{\rm D}^{*}\right)_{}^{-1} M_{\rm R}\right]_{}^{-1}\tilde{H}_{\rm D}^{}\left[M_{\rm R}^{\dagger}\left(\tilde{H}_{\rm D}^{*}\right)_{}^{-1} M_{\rm R}\right]_{}^{-1}\right\}\nonumber\\
&=&\left[I_{004}^{}\left(I_{020}^2-I_{040}^{}\right)_{}^2+ I_{002}^2\left(3I_{020}^2-I_{040}^{}\right)\left(I_{020}^2+ I_{040}^{}\right)-4\left(2I_{020}^{}I_{022}^{}-I_{042}^{}\right) \right.\nonumber\\
&&\left. \times \left(2I_{002}^{}I_{020}^2-2I_{020}^{}I_{022}^{}+I_{042}^{}\right)\right] /\left(I_{002}^2-I_{004}^{}\right)_{}^2\;.
\end{eqnarray}

From the perspective of effective theories, the matching conditions for the flavor invariants in Eqs.~(\ref{eq:match Hnu})-(\ref{eq:match Hnu2}) are valid at the seesaw scale. Below the seesaw scale where the RH neutrinos have been integrated out, the running behaviors of the flavor invariants are governed by those of $M_\nu^{}$ and $M_l^{}$. To obtain the values of flavor invariants at the electroweak scale, one needs to solve the RGEs of $\left\{I_1^{},...,I_{34}^{}\right\}$, which have been calculated in Ref.~\cite{Wang2021}. Therefore, once the UV-complete model is specified, the matching conditions will be used to determine the initial values of the flavor invariants in the effective theory. With the RGEs of the basic invariants in Table~\ref{table:MSM} in the full theory and the matching conditions at the seesaw scale, we have given a complete description of the running behaviors of the invariants in the low-energy effective theory.

Using the Casas-Ibarra parametrization in Eq.~(\ref{eq:CI parametrization}), it is possible to extract all the physical parameters from the basic invariants in Table~\ref{table:MSM}. This goal can be achieved as follows. First of all, the lightest active neutrino is massless in the MSM, so we have $I_2^{}=m_2^2+m_3^2$ and $I_5^{}=m_2^4+m_3^4$, from which one can extract the masses of the active neutrinos
\begin{eqnarray}
\label{eq:extract mi}
m_{2}^{2}=\frac{1}{2}\left(I_2^{} - \sqrt{2I_5^{}-I_2^2}\right)\;, \quad m_{3}^{2}=\frac{1}{2}\left(I_2^{} + \sqrt{2I_5^{}-I_2^2}\right)\;,
\end{eqnarray}
where $I_2^{}$ and $I_5^{}$ are given by Eqs.~(\ref{eq:match Hnu})-(\ref{eq:match Hnu2}). The masses of two heavy Majorana neutrinos are determined by $I_{002}^{}\equiv {\rm Tr}\left(H_{\rm R}^{}\right)=M_1^2+M_2^2$ and $I_{004}^{}\equiv {\rm Tr}\left(H_{\rm R}^{2}\right)=M_1^4+M_2^4$, i.e.,
\begin{eqnarray}
\label{eq:extract Mi}
M_{1}^2=\frac{1}{2}\left(I_{002}^{} - \sqrt{2I_{004}^{}-I_{002}^2}\right)\;, \quad M_{2}^2=\frac{1}{2}\left(I_{002}^{} + \sqrt{2I_{004}^{}-I_{002}^2}\right)\;.
\end{eqnarray}
As for the masses of charged-leptons, one can use $I_{200}^{}\equiv {\rm Tr}\left(H_l^{}\right)=m_e^2+m_\mu^2+m_\tau^2$, $I_{400}^{}\equiv {\rm Tr}\left(H_l^{2}\right)=m_e^4+m_\mu^4+m_\tau^4$ and $I_{600}^{}\equiv {\rm Tr}\left(H_l^{3}\right)=m_e^6+m_\mu^6+m_\tau^6$, leading to
\begin{eqnarray}
m_e^2 = \frac{I_{200}^3 - 3I^{}_{200} I^{}_{400} + 2I^{}_{600}}{3\left(I_{200}^2-I^{}_{400}\right)} \;,\quad
m_\mu ^2 = \frac{I_{200}^2-I^{}_{400}}{2 I_{600}^{1/3}}\;,\quad
m_\tau ^2 = I_{600}^{1/3}\;,
\end{eqnarray}
where the hierarchical limit $m_\tau^{}\gg m_\mu^{}\gg m_e^{}$ has been applied. In addition, the real and imaginary parts of the complex parameter $z\equiv x+{\rm i}y$ can be solved from the following identities
\begin{eqnarray*}
I_{020}^{}&\equiv& {\rm Tr}\left(H_{\rm D}^{}\right)=
\frac{1}{2}\left[\left(M_1^{}-M_2^{}\right)\left(m_2^{}-m_3^{}\right)\cos \left(2x\right)+\left(M_1^{}+M_2^{}\right)\left(m_2^{}+m_3^{}\right)\cosh \left(2y\right) \right]\;,\\
I_{022}^{}&\equiv& {\rm Tr}\left(\tilde{H}_{\rm D}^{}H_{\rm R}^{}\right)=
\frac{1}{2}\left[\left(M_1^{3}-M_2^{3}\right)\left(m_2^{}-m_3^{}\right) \cos\left(2x\right)+\left(M_1^{3}+M_2^{3}\right)\left(m_2^{}+m_3^{}\right)\cosh\left(2y\right) \right]\;.
\end{eqnarray*}
More explicitly, we have
\begin{eqnarray}
\label{eq:extract z}
\cos\left(2x\right) &=& \frac{I^{}_{022}-I^{}_{020} \left(M_1^2 - M^{}_1 M^{}_2 + M_2^2\right)}{M^{}_1 M^{}_2 \left(M^{}_1 - M^{}_2\right) \left(m^{}_2 - m^{}_3\right)} \;, \nonumber\\
\cosh\left(2y\right) &=& \frac{-I^{}_{022} + I^{}_{020} \left(M_1^2 + M^{}_1 M^{}_2 + M_2^2\right)}{M^{}_1 M^{}_2 \left(M^{}_1 + M^{}_2\right) \left(m^{}_2 + m^{}_3\right)}\;,
\end{eqnarray}
where the masses of RH neutrinos are assumed to be non-degenerate, i.e., $M_1^{}\neq M_2^{}$. The latest global-fit analysis of neutrino oscillation data~\cite{Esteban:2020cvm} indicates $\Delta m^2_{31} \equiv m^2_3 - m^2_1 \approx 2.51\times 10^{-3}~{\rm eV}^2$ and $\Delta m^2_{21} \equiv m^2_2 - m^2_1 \approx 7.42\times 10^{-5}~{\rm eV}^2$, from which one can verify that $m^{}_2 \neq m^{}_3$ in the MSM with $m^{}_1 = 0$. Therefore, the first identity in Eq.~(\ref{eq:extract z}) is valid. Substituting Eqs.~(\ref{eq:extract mi}) and (\ref{eq:extract Mi}) into Eq.~(\ref{eq:extract z}) and doing some arithmetical computations, one can further reexpress the results completely in terms of the flavor invariants, namely,
\begin{eqnarray}
\label{eq:extract z2}
\cos\left(2x\right)&=&
\frac{\sqrt{2}I_{020}^{}\sqrt{I_{002}^2-I_{004}}-2I_{002}I_{020}+2I_{022}}{\sqrt{I_{002}^2-I_{004}}\left(
\sqrt{I_{002}-\sqrt{2I_{004}-I_{002}^2}}-\sqrt{I_{002}+\sqrt{2I_{004}-I_{002}^2}}\right)}\nonumber\\
&&\times \frac{\sqrt{2}}{\sqrt{I_2^{}-\sqrt{2I_5-I_2^2}}-\sqrt{I_2+\sqrt{2I_5-I_2^2}}}\;,\nonumber\\
\cosh\left(2y\right)&=&
\frac{\sqrt{2}I_{020}^{}\sqrt{I_{002}^2-I_{004}}+2I_{002}I_{020}-2I_{022}}{\sqrt{I_{002}^2-I_{004}}\left(
\sqrt{I_{002}-\sqrt{2I_{004}-I_{002}^2}}+\sqrt{I_{002}+\sqrt{2I_{004}-I_{002}^2}}\right)}\nonumber\\
&& \times \frac{\sqrt{2}}{\sqrt{I_2^{}-\sqrt{2I_5-I_2^2}}+\sqrt{I_2+\sqrt{2I_5-I_2^2}}}\;,
\end{eqnarray}
where the explicit expressions of flavor invariants $I^{}_2$ and $I^{}_5$ can be found in Eqs.~(\ref{eq:match Hnu})-(\ref{eq:match Hnu2}).
Finally, the flavor mixing angles $\left\{\theta_{12}^{},\theta_{13}^{},\theta_{23}^{}\right\}$ and CP phases $\left\{\delta,\sigma\right\}$ in the PMNS matrix can be extracted from $\left\{I_1^{},I_2^{},...I_{34}^{} \right\}$ as shown in Ref.~\cite{Wang2021}, which in turn can be recast into the rational functions of the basic invariants in Table~\ref{table:MSM} by virtue of Eq.~(\ref{eq:matching identity}). Therefore, we have successfully extracted all the physical parameters from the basic invariants in the MSM as promised. Complemented with the RGEs of the flavor invariants, the above relations offer a basis-independent way to describe the running behaviors of physical parameters.

\subsection{Conditions for CP Conservation}
In Ref.~\cite{Yu2020PRD} we have found three sufficient and necessary conditions for CP conservation in the leptonic sector in the MSM. As a simple application of the flavor invariants, we show that those conditions can be written in a basis-independent form with only the basic invariants in Table~\ref{table:MSM}.

In the assumption that the masses of heavy Majorana neutrinos are not degenerate, there are totally three CP-violating phases in the theory. In the present case, we choose three CP-violating parameters to be $y$, $\delta$ and $\sigma$, implying that three CP-odd invariants are needed to eliminate them. First, the lowest-order CP-odd invariant $I_{044}^{}$ can be used to get rid of $y$ via
\begin{eqnarray}
\label{eq:CP odd y}
I_{044}^{}\equiv {\rm Tr}\left(\left[H_{\rm R}^{},\tilde{H}_{\rm D}^{}\right]G_{\rm DR}\right)={\rm i} M_1^2 M_2^2 \left(M_1^2-M_2^2\right)\left(m_2^2-m_3^2\right)\sin\left(2x\right)\sinh\left(2y\right)\;,
\end{eqnarray}
the vanishing of which results in $y=0$.\footnote{The special values of $x=0$ or $\pi/2$ also lead to the vanishing of $I_{044}^{}$. In these two cases, the orthogonal matrix $R$ will be reduced to a rotation matrix with a purely-imaginary rotation angle. However, it can be shown that the phase of $\pi/2$ in $R$ is actually unphysical and can be absorbed into other phases~\cite{Yu2020PRD}. Therefore, $I_{044}^{}=0$ can indeed eliminate one CP phase.} To separate the Dirac-type CP phase $\delta$ from the Majorana-type CP phase $\sigma$, we recall the Jarlskog-type invariant in the low-energy effective theory
\begin{eqnarray}
I_{25}^{}\equiv \frac{1}{3}{\rm Tr}\left(\left[H_l^{},H_\nu^{}\right]_{}^3 \right)=2{\rm i}m_2^{2}m_3^{2}\left(m_3^2-m_2^2\right)\Delta_{e\mu}^{}\Delta_{\mu\tau}^{}\Delta_{\tau e}^{} s_{12}^{}c_{12}^{}s_{23}^{}c_{23}^{}s_{13}^{}c_{13}^{2}\sin\delta\;,
\end{eqnarray}
where $\Delta_{\alpha \beta}^{}\equiv m_\alpha^2-m_\beta^2$ (for $\alpha,\beta=e, \mu, \tau$), $c_{ij}^{}\equiv \cos \theta_{ij}^{}$ and $s_{ij}^{}\equiv \sin \theta_{ij}^{}$ (for $ij=12,13,23$). Thus the vanishing of $I_{25}^{}$ enforces $\delta=0$. Our task is to express $I_{25}^{}$ in the form of those basic invariants in Table~\ref{table:MSM}. Replacing $M_\nu$ with $-M_{\rm D}^{}M_{\rm R}^{-1}M_{\rm D}^{\rm T}$  in $I_{25}^{}$ and taking advantage of Eq.~(\ref{eq:matching identity}) repeatedly, one obtains
\begin{eqnarray}
	\label{eq:I25 mid}
I_{25}^{}&=&\frac{-16\, {\rm i}}{\left(I_{002}^2-I_{004}\right)^3}{\rm Im Tr}
\left\{G_{l\rm D}^{(2)}
\left[I_{020}^{}H_{\rm R}^{}-G_{\rm DR}-\left(I_{020}^{}I_{002}^{}-I_{022}^{}\right){\bf 1}_2^{}\right]\tilde{H}_{\rm D}^{}\right.\nonumber\\
&&\left.\times \left[I_{020}^{}H_{\rm R}^{}-G_{\rm DR}-\left(I_{020}^{}I_{002}^{}-I_{022}^{}\right){\bf 1}_2^{}\right] G_{l\rm D}^{}\left[I_{020}^{}H_{\rm R}^{}-G_{\rm DR}-\left(I_{020}^{}I_{002}^{}-I_{022}^{}\right){\bf 1}_2^{}\right]
\right\}\;.
\end{eqnarray}
In order to decompose the flavor invariants in Eq.~(\ref{eq:I25 mid}) into the polynomials of the basic invariants, the Cayley-Hamilton theorem should be used. The final result turns out to be lengthy
\begin{eqnarray}
\label{eq:CP odd delta}
I_{25}^{}={\mathscr I}/\left(I_{002}^2-I_{004}^{}\right)_{}^3\;,
\end{eqnarray}
with
{\allowdisplaybreaks
\begin{eqnarray*}
{\mathscr I} &=&-2I_{020}^4 \left(3 I_{002}^2-2I_{004}^{}\right)I_{642}^{(1)}+I_{020}^{3}\left[5I_{002}^3 I_{660}^{}-I_{002}^2 \left(6I_{242}^{(2)}I_{420}+6 I_{220}^{}I_{442}^{(3)}-5 I_{662}^{(1)}+5 I_{662}^{(2)}\right)
\right.\nonumber\\
&& \left. + I_{002}^{}\left(20 I_{242}^{(2)} I_{422} + 20 I_{222} I_{442}^{(3)} + 12 I_{022}^{} I_{642}^{(1)} - 19 I_{004}^{} I_{660}^{}\right)+I_{004}^{} \left(4 I_{242}^{(2)} I_{420}^{} + 4 I_{220}^{} I_{442}^{(3)} - I_{662}^{(1)}\right.\right.\nonumber\\
&& \left.\left. + I_{662}^{(2)}\right)\right] + I_{020}^2\left\{
I_{002}^2 \left(7 I_{262}^{} I_{420}^{} - 5 I_{242}^{(2)} I_{440}^{} - 5 I_{240}^{} I_{442}^{(3)} + 7 I_{220}^{} I_{462}^{(2)} - 5 I_{040}^{} I_{642}^{(1)} - 6 I_{022}^{} I_{660}^{}\right)\right.\nonumber\\
&& \left. -2 I_{002}^{}\left[I_{022}^{} \left(6 I_{242}^{(2)} I_{420}^{} + 6 I_{220}^{} I_{442}^{(3)} + 9 I_{662}^{(1)} - 9 I_{662}^{(2)}\right)
-2\left(I_{240}^{} I_{444}^{(2)}+I_{244}^{}I_{440}^{}+2I_{042}^{}I_{642}^{(1)} \right)\right.\right.\nonumber\\
&& \left.\left. +9\left(I_{222}^{} I_{462}^{(2)}+ I_{262}^{} I_{422}^{}\right)-7\left(I_{242}^{(1)} I_{442}^{(3)}+I_{242}^{(2)} I_{442}^{(2)} \right)\right]-6 I_{022}^{} \left(I_{244}^{} I_{420}^{} + 5 I_{242}^{(2)} I_{422}^{} + 5 I_{222}^{} I_{442}^{(3)} \right.\right.\nonumber\\
&&\left.\left. + I_{220}^{} I_{444}^{(2)}\right) - 8 I_{022}^2 I_{642}^{(1)}+I_{004}^{} \left(I_{262}^{} I_{420}^{} - 7 I_{242}^{(2)} I_{440}^{} - 7 I_{240}^{} I_{442}^{(3)} + I_{220}^{} I_{462}^{(2)} - 7 I_{040}^{} I_{642}^{(1)}\right.\right.\nonumber\\
&&\left.\left. + 38 I_{022}^{} I_{660}^{}\right)+4 I_{044}^{} \left(I_{222}^{} I_{420}^{} - I_{220}^{} I_{422}^{}\right)
\right\}+I_{020}^{} \left\{9 I_{002}^3 I_{040}^{} I_{660}+I_{002}^2\left[
6 \left(I_{262}^{} I_{440}^{} + I_{240} I_{462}^{(2)}\right.\right.\right.\nonumber\\
&& \left.\left.\left. - 2 I_{042}^{} I_{660}^{}\right) +
 I_{040}^{} \left(2 I_{242}^{(2)} I_{420}^{} + 2 I_{220}^{} I_{442}^{(3)} + 5 I_{662}^{(1)} - 5 I_{662}^{(2)}\right)\right]+I_{002}^{}\left[
3 I_{040}^{} \left(2 I_{242}^{(2)} I_{422}^{}\right.\right.\right.\nonumber\\
&& \left.\left.\left.  + 2 I_{222}^{} I_{442}^{(3)}- 3 I_{004}^{} I_{660}^{}\right)-6 I_{022}^{} \left(I_{262}^{} I_{420}^{} - 3 I_{242}^{(2)} I_{440}^{} - 3 I_{240}^{} I_{442}^{(3)} + I_{220}^{} I_{462}^{(2)} - 4 I_{040} I_{642}^{(1)}\right)\right.\right.\nonumber\\
&& \left.\left. -2 I_{042}^{}\left(I_{242}^{(2)} I_{420}^{} + I_{220}^{} I_{442}^{(3)} + 4 I_{662}^{(1)} - 4 I_{662}^{(2)}\right)-8 \left(I_{262}^{} I_{442}^{(2)} + I_{242}^{(1)} I_{462}^{(2)}\right)\right]+ 4 I_{220}^{} \left( I_{044}^{} I_{442}^{(2)}\right.\right.\nonumber\\
&& \left.\left. +  I_{042}^{} I_{444}^{(2)}\right)+24 I_{022}^2 \left(I_{242}^{(2)} I_{420}^{} + I_{220}^{} I_{442}^{(3)} + I_{662}^{(1)} - I_{662}^{(2)}\right)+2 I_{022}^{} \left(13 I_{262}^{} I_{422}^{}+ 13 I_{222}^{} I_{462}^{(2)}\right.\right.\nonumber\\
&& \left.\left. - 15 I_{242}^{(2)} I_{442}^{(2)} - 15 I_{242}^{(1)} I_{442}^{(3)}  - 16 I_{042}^{} I_{642}^{(1)}\right)+I_{004}^{} \left[4 I_{042}^{} I_{660}^{} - I_{040}^{} \left(2 I_{242}^{(2)} I_{420}^{} + 2 I_{220}^{} I_{442}^{(3)} - I_{662}^{(1)}\right.\right.\right.\nonumber\\
&& \left.\left.\left. + I_{662}^{(2)}\right)\right]+4 I_{042}^{} \left( I_{244}^{} I_{420}^{} - I_{242}^{(2)} I_{422}^{} - I_{222}^{} I_{442}^{(3)}\right)-4 I_{044}^{} I_{242}^{(1)} I_{420}^{}\right\}-2 I_{022}^2 \left[I_{262}^{}I_{420}^{} + I_{220}^{} I_{462}^{(2)}\right.\nonumber\\
&& \left. - 5 \left(I_{242}^{(2)} I_{440}^{} + I_{240}^{} I_{442}^{(3)}\right)\right]-2 I_{022}^{}\left\{I_{002}\left[5 \left(I_{262}^{} I_{440}^{} + I_{240}^{} I_{462}^{(2)} + I_{040}^{}I_{662}^{(1)} - I_{040}^{}I_{662}^{(2)}\right)\right.\right.\nonumber\\
&& \left.\left. - 6 I_{042}^{} I_{660}^{}\right]+6 I_{002}^2 I_{040}^{} I_{660}^{}+I_{040}^{}\left(5 I_{242}^{(2)} I_{422}^{} + 5 I_{222}^{} I_{442}^{(3)} - 6 I_{004}^{} I_{660}^{}\right)-I_{042}^{}\left(I_{242}^{(2)} I_{420}^{}\right.\right.\nonumber\\
&& \left.\left. + I_{220}^{} I_{442}^{(3)}+ 2 I_{662}^{(1)} - 2 I_{662}^{(2)}\right)-2 \left(I_{262}^{} I_{442}^{(2)} + I_{242}^{(1)} I_{462}^{(2)}\right)\right\}-12 I_{022}^3 I_{660}^{}-I_{040}^{}\left(I_{002}^2 - I_{004}\right)\nonumber\\
&&\times \left(I_{262}^{} I_{420}^{} - I_{242}^{(2)} I_{440}^{} - I_{240}^{} I_{442}^{(3)} + I_{220}^{} I_{462}^{(2)} - I_{040}^{} I_{642}^{(1)}\right)
\;,
\end{eqnarray*}}
which though tedious is straightforward to verify. Therefore, we have written $I^{}_{25}$ as the rational function of the basic invariants in Table~\ref{table:MSM} as expected. The last step is, after using Eqs.~(\ref{eq:CP odd y}) and (\ref{eq:CP odd delta}) to eliminate $y$ and $\delta$, to find another CP-odd invariant $I_{242}^{(2)}$, namely,
\begin{eqnarray}
\label{eq:CP odd sigma}
I_{242}^{(2)}\equiv {\rm Tr}\left(\left[H_{\rm R}^{},\tilde{H}_{\rm D}^{} \right]G_{l\rm D}^{}\right)\xlongequal{y=\delta=0}&&\frac{{\rm i}}{2}M_1^{}M_2^{}\left(M_2^2-M_1^2\right)\left(m_2^{}-m_3^{}\right)\sqrt{m_2^{}m_3^{}}c_{13}^{}\sin 2x\left[\left(\Delta_{e\mu}^{}
\right.\right.\nonumber\\
&&\left.\left. +\Delta_{e\tau}^{} +\Delta_{\mu\tau}^{}\cos 2\theta_{23}^{}\right)s_{12}^{}s_{13}^{} +\Delta_{\mu \tau}^{}c_{12}^{}\sin 2\theta_{23}^{}\right]\sin\sigma\;,
\end{eqnarray}
which is proportional to $\sin\sigma$. Obviously, the vanishing of $I_{242}^{(2)}$ ultimately removes all the CP phases in the theory.

To conclude, the vanishing of Eqs.~(\ref{eq:CP odd y}), (\ref{eq:CP odd delta}) and (\ref{eq:CP odd sigma}) in the MSM gives the sufficient and necessary conditions for CP conservation in the leptonic sector. These conditions, put in the form of only the basic invariants in Table~\ref{table:MSM}, are dependent on neither the chosen flavor basis nor the parametrization of $M_{\rm D}^{}$.

\subsection{CP Asymmetries in Leptogenesis}
Apart from naturally generating tiny Majorana masses of neutrinos, the seesaw model provides an elegant possibility to explain the matter-antimatter asymmetry of our Universe by the leptogenesis mechanism~\cite{Fukugita1986}, where the lepton number asymmetries in the CP-violating and out-of-equilibrium decays of heavy Majorana neutrinos can be converted into the baryon number asymmetry via the sphaleron processes. The links between the leptonic CP violation and the flavor invariants in the leptogenesis were first discussed in Ref.~\cite{Pilaftsis1997} and subsequently examined in several other works~\cite{Branco2001,CIP2006}. In this subsection, we show that the CP asymmetries in the decays of heavy Majorana neutrinos can be expressed in a simple form with only the basic invariants.

For simplicity, we consider the scenario where the one-flavor approximation for leptogenesis is working well. Therefore, only the CP asymmetries summed over lepton flavors are relevant
\begin{eqnarray}
\epsilon_i^{}\equiv \frac{\sum\nolimits_{\alpha} \left[\Gamma\left(\nu_{i\rm R}^{}\rightarrow \ell_\alpha+H\right)-\Gamma\left(\nu_{i\rm R}\rightarrow \overline{\ell_\alpha}+\overline{H}\right)\right]}{\sum\nolimits_{\alpha} \left[\Gamma\left(\nu_{i\rm R}^{}\rightarrow \ell_\alpha+H\right)+\Gamma\left(\nu_{i\rm R}\rightarrow \overline{\ell_\alpha}+\overline{H}\right)\right]}\;,
\end{eqnarray}
where $\Gamma \left(\nu_{i\rm R}^{}\rightarrow \ell_\alpha^{}+H\right)$ and $\Gamma\left(\nu_{i\rm R}^{}\rightarrow \overline{\ell_\alpha^{}}+\overline{H}\right)$ stand respectively for the decay rate of $\nu_{i\rm R}^{}\rightarrow \ell_\alpha^{}+H$ and that of $\nu_{i\rm R}^{}\rightarrow \overline{\ell_\alpha^{}}+\overline{H}$, with $\overline{\ell_\alpha^{}}$ (for $\alpha=e,\mu,\tau$) and $\overline{H}$ being the CP-conjugated states of the lepton and Higgs doublets. In the MSM, the CP asymmetries arise from the interference between the tree- and one-loop-level decay amplitudes and are given by~\cite{XZ2011}
\begin{eqnarray}
\label{eq:CP asym}
\epsilon_i^{}=\frac{1}{4\pi v^2\left(\tilde{H}_{\rm D}\right)_{ii} }\sum_{j\neq i}^{}{\rm Im}\left[\left(\tilde{H}_{\rm D}\right)_{ij}^2 \right]{\cal F}\left(\frac{M_j^2}{M_i^2}\right)\quad ({\rm for}\; i=1,2)\;,
\end{eqnarray}
where the loop function is defined as
\begin{eqnarray*}
{\cal F}\left(x\right)\equiv \sqrt{x}\left[\frac{2-x}{1-x}+\left(1+x\right)\ln\left(\frac{x}{1+x}\right)\right]\;.
\end{eqnarray*}
First, we insert the parametrization of $M_{\rm D}^{}$ in Eq.~(\ref{eq:CI parametrization}) into $\tilde{H}^{}_{\rm D}$ in Eq.~(\ref{eq:CP asym}) and thus obtain
\begin{eqnarray}
\label{eq:CP asym parame}
\epsilon_1^{}=\frac{1}{4\pi v^2}\frac{M_2\left(m_2^2-m_3^2\right)\sin\left(2x\right)\sinh\left(2y\right)}{\left(m_2-m_3\right)\cos\left(2x\right)+\left(m_2+m_3\right)\cosh\left(2y\right)}
{\cal F}\left(\frac{M_2^2}{M_1^2}\right)\;,\nonumber\\
\epsilon_2^{}=\frac{1}{4\pi v^2}\frac{M_1\left(m_2^2-m_3^2\right)\sin\left(2x\right)\sinh\left(2y\right)}{\left(m_2-m_3\right)\cos\left(2x\right)-\left(m_2+m_3\right)\cosh\left(2y\right)}
{\cal F}\left(\frac{M_1^2}{M_2^2}\right)\;.
\end{eqnarray}
Then, recalling the extraction of the physical parameters in Eqs.~(\ref{eq:extract mi}) and (\ref{eq:extract z}) and substituting them into Eq.~(\ref{eq:CP asym parame}), we get the CP asymmetries in the form of the basic invariants
\begin{eqnarray}
\label{eq:CP asym inv}
\epsilon_1^{}=\frac{-{\rm i}\, I_{044}^{}}{8\pi v^2 M_1 M_2\left(I_{022}^{}-M_2^2I_{020}\right)}{\cal F}\left(\frac{M_2^2}{M_1^2}\right)\;,\nonumber\\
\epsilon_2^{}=\frac{-{\rm i}\, I_{044}^{}}{8\pi v^2 M_1 M_2\left(I_{022}^{}-M_1^2I_{020}\right)}{\cal F}\left(\frac{M_1^2}{M_2^2}\right)\;,
\end{eqnarray}
where $M_1^{}$ and $M_2^{}$ are given by Eq.~(\ref{eq:extract Mi}). Note that the imaginary unit in the expressions of $\epsilon_i^{}$  (for $i=1,2$) in Eq.~(\ref{eq:CP asym inv}) will be canceled out by the imaginary unit in the flavor invariant $I_{044}^{}$ (cf. Eq.~(\ref{eq:CP odd y})), ensuring that the CP asymmetries are real. In particular, when the mass spectrum of heavy Majorana neutrinos is hierarchical with $M_2^{}\gg M_1^{}$, only the CP asymmetry $\epsilon^{}_1$ from the lighter one is relevant. The formula of the CP asymmetry in this case is greatly simplified
\begin{eqnarray}
\epsilon_1^{}\approx\frac{3\,{\rm i}}{16\pi v^2}\frac{I^{}_{044}}{I_{002}\left(I^{}_{022}-I_{002}^{}I^{}_{020}\right)}\;,
\end{eqnarray}
where it is evident that only the basic flavor invariants are involved. As we have mentioned in the previous subsection, the vanishing of $I^{}_{044}$ eliminates one CP-violating parameter, so the CP asymmetry $\epsilon^{}_1$ here vanishes accordingly.

\section{Summary}
\label{sec:summary}
In this paper, we investigate the flavor invariants in the minimal seesaw model by using the Hibert series and the plethystic logarithm. Complementary to the previous work ~\cite{Wang2021}, in which the flavor invariants in the low-energy effective theory have been studied, the explicit construction of flavor invariants in a complete seesaw model is accomplished. Our main results and conclusions are summarized below.

First, the Hilbert series for the flavor invariants in the minimal seesaw model has been computed for the first time, as shown in Eqs.~(\ref{eq:multi HS MSM}) and (\ref{eq:ungraded HS MSM}). Then, with the help of the Hilbert series and the plethystic logarithm we explicitly construct all the basic invariants, which have been listed in Table~\ref{table:MSM}. We find that there are in total 38 basic invariants, among which 18 invariants are CP-odd and the others are CP-even. Any flavor invariants in the minimal seesaw model constructed from the matrix polynomials of $M_l^{}$, $M_{\rm D}^{}$ and $M_{\rm R}^{}$ can be decomposed into the polynomials of these 38 basic invariants. All the physical parameters can also be extracted from the basic invariants.

Furthermore, we investigate the relationship between the flavor invariants in a UV-complete model and those in the corresponding low-energy effective theory. As explained in Sec.~\ref{subsec:relation}, any flavor invariants at the low-energy scale can be written as the rational functions of those at the high-energy scale. These rational functions serve as the matching conditions and supply a UV-complete description of the running behaviors of the flavor invariants in the effective theory. We have also discussed some phenomenological applications of the flavor invariants in the minimal seesaw model. We reexamine the sufficient and necessary conditions for CP conservation in the leptonic sector as well as the CP asymmetries in the decays of heavy Majorana neutrinos. All these physically interesting quantities have been successfully expressed in terms of only the basic invariants in Table~\ref{table:MSM}.

Thus far it remains unknown how neutrino masses and lepton flavor mixing are generated, and different theories at the high-energy scale may lead to the same low-energy effective theory. In this sense, it is necessary to study different complete theories that exhibit distinct flavor structures and invariant rings. Our formalism for the minimal type-I seesaw model can be easily generalized to other complete theories, such as the minimal type-(I+II) seesaw model, which extends the SM with one RH neutrino and one scalar triplet~\cite{Gu2006}.
As we have seen, the invariant theory and the flavor invariants are extremely useful in studying the flavor structures of fermions as well as the CP violation in the quark and leptonic sector. Moreover, they also provide a novel way to establish the relationship between the complete theories and their low-energy effective counterparts. The applications of invariant theory to flavor physics are in the very early stage, and more dedicated studies are desired.

\section*{Acknowledgements}
This work was supported in part by the National Natural Science Foundation of China under grant No. 11775232 and No. 11835013, by the Key Research Program of the Chinese Academy of Sciences under grant No. XDPB15, and by the CAS Center for Excellence in Particle Physics.


\begin{thebibliography}{99}

\bibitem{PDG2020}
P.~A.~Zyla \textit{et al.} [Particle Data Group],
``Review of Particle Physics,''
PTEP \textbf{2020}, no.8, 083C01 (2020).

\bibitem{Xing2020}
Z.~z.~Xing,
``Flavor structures of charged fermions and massive neutrinos,''
Phys. Rept. \textbf{854}, 1-147 (2020)
[arXiv:1909.09610 [hep-ph]].

\bibitem{Minkowski1977}
P.~Minkowski,
``$\mu \to e\gamma$ at a Rate of One Out of $10^{9}$ Muon Decays?,''
Phys. Lett. B \textbf{67} (1977), 421-428

\bibitem{GellMann1979}
M.~Gell-Mann, P.~Ramond and R.~Slansky,
``Complex Spinors and Unified Theories,''
Conf. Proc. C \textbf{790927} (1979), 315-321
[arXiv:1306.4669 [hep-th]].

\bibitem{Yanagida1980}
T.~Yanagida,
``Horizontal Symmetry and Masses of Neutrinos,''
Prog. Theor. Phys. \textbf{64} (1980), 1103

\bibitem{Glashow1980}
S.~L.~Glashow,
``The Future of Elementary Particle Physics,''
NATO Sci. Ser. B \textbf{61} (1980), 687

\bibitem{Mohapatra1980}
R.~N.~Mohapatra and G.~Senjanovic,
``Neutrino Mass and Spontaneous Parity Nonconservation,''
Phys. Rev. Lett. \textbf{44} (1980), 912

\bibitem{Majorana1937}
E.~Majorana,
``Teoria simmetrica dell\textquoteright{}elettrone e del positrone,''
Nuovo Cim. \textbf{14}, 171-184 (1937).

\bibitem{Racah1937}
G.~Racah,
``On the symmetry of particle and antiparticle,''
Nuovo Cim. \textbf{14}, 322-328 (1937).


\bibitem{Jarlskog1985a}
C.~Jarlskog,
``Commutator of the Quark Mass Matrices in the Standard Electroweak Model and a Measure of Maximal CP Violation,''
Phys. Rev. Lett. \textbf{55}, 1039 (1985)

\bibitem{Jarlskog1985b}
C.~Jarlskog,
``A Basis Independent Formulation of the Connection Between Quark Mass Matrices, CP Violation and Experiment,''
Z. Phys. C \textbf{29}, 491-497 (1985)


\bibitem{Wu1986}
D.~d.~Wu,
``The Rephasing Invariants and CP,''
Phys. Rev. D \textbf{33}, 860 (1986)


\bibitem{Branco1986quark}
J.~Bernabeu, G.~C.~Branco and M.~Gronau,
``CP Restrictions on Quark Mass Matrices,''
Phys. Lett. B \textbf{169}, 243-247 (1986).

\bibitem{Branco1986lepton}
G.~C.~Branco, L.~Lavoura and M.~N.~Rebelo,
``Majorana Neutrinos and CP Violation in the Leptonic Sector,''
Phys. Lett. B \textbf{180}, 264-268 (1986).

\bibitem{Yu2019PLB}
B.~Yu and S.~Zhou,
``The number of sufficient and necessary conditions for CP conservation with Majorana neutrinos: three or four?,''
Phys. Lett. B \textbf{800}, 135085 (2020)
[arXiv:1908.09306 [hep-ph]].


\bibitem{Yu2020ICHEP}
B.~Yu and S.~Zhou,
``Weak-basis invariants and CP conservation in the leptonic sector with Majorana neutrinos,''
[arXiv:2010.08758 [hep-ph]].


\bibitem{Yu2020PRD}
B.~Yu and S.~Zhou,
``Sufficient and Necessary Conditions for CP Conservation in the Case of Degenerate Majorana Neutrino Masses,''
Phys. Rev. D \textbf{103}, no.3, 035017 (2021)
[arXiv:2009.12347 [hep-ph]].


\bibitem{Wang2021}
Y.~Wang, B.~Yu and S.~Zhou,
``Flavor Invariants and Renormalization-group Equations in the Leptonic Sector with Massive Majorana Neutrinos,''
[arXiv:2107.06274 [hep-ph]].

\bibitem{Sturmfels2008}
B.~Sturmfels
``Algorithms in Invariant Theory,''
  Springer-Verlag, Wien (2008).

\bibitem{DK2015}
H.~Derksen, G.~Kemper, V.~L.~Popov and N.~A’~Campo,
``Computational invariant theory,''
 Springer-Verlag, Berlin Heidelberg (2015).


\bibitem{JM2009}
  E.~E.~Jenkins and A.~V.~Manohar,
  ``Algebraic Structure of Lepton and Quark Flavor Invariants and CP Violation,''
  JHEP {\bf 0910}, 094 (2009)
  [arXiv:0907.4763 [hep-ph]].


\bibitem{HJMT2011}
A.~Hanany, E.~E.~Jenkins, A.~V.~Manohar and G.~Torri,
``Hilbert Series for Flavor Invariants of the Standard Model,''
JHEP \textbf{03}, 096 (2011)
[arXiv:1010.3161 [hep-ph]].


\bibitem{Kleppe1995}
  A.~Kleppe,
  ``Extending the standard model with two right-handed neutrinos,''
in {\it Neutrino physics. Proceedings of 3rd Tallinn Symposium}, Lohusalu, Estonia, October 8-11, 1995, page 118-125.

\bibitem{Ma1998}
E.~Ma, D.~P.~Roy and U.~Sarkar,
``A Seesaw model for atmospheric and solar neutrino oscillations,''
Phys. Lett. B \textbf{444}, 391-396 (1998)
[arXiv:hep-ph/9810309].


\bibitem{King1999}
S.~F.~King,
``Large mixing angle MSW and atmospheric neutrinos from single right-handed neutrino dominance and U(1) family symmetry,''
Nucl. Phys. B \textbf{576}, 85-105 (2000)
[arXiv:hep-ph/9912492].


\bibitem{Lavoura2000}
L.~Lavoura and W.~Grimus,
``Seesaw model with softly broken L(e) - L(muon) - L(tau),''
JHEP \textbf{09}, 007 (2000)
[arXiv:hep-ph/0008020 [hep-ph]].


\bibitem{King2002}
S.~F.~King,
``Constructing the large mixing angle MNS matrix in seesaw models with right-handed neutrino dominance,''
JHEP \textbf{09}, 011 (2002)
[arXiv:hep-ph/0204360].

\bibitem{Frampton2002}
P.~H.~Frampton, S.~L.~Glashow and T.~Yanagida,
``Cosmological sign of neutrino CP violation,''
Phys. Lett. B \textbf{548}, 119-121 (2002)
[arXiv:hep-ph/0208157].

\bibitem{Guo2006}
W.~l.~Guo, Z.~z.~Xing and S.~Zhou,
``Neutrino Masses, Lepton Flavor Mixing and Leptogenesis in the Minimal Seesaw Model,''
Int. J. Mod. Phys. E \textbf{16}, 1-50 (2007)
[arXiv:hep-ph/0612033].

\bibitem{XZ2020}
Z.~z.~Xing and Z.~h.~Zhao,
``The minimal seesaw and leptogenesis models,''
Rept. Prog. Phys. \textbf{84}, no.6, 066201 (2021)
[arXiv:2008.12090 [hep-ph]].

\bibitem{Fukugita1986}
M.~Fukugita and T.~Yanagida,
``Baryogenesis Without Grand Unification,''
Phys. Lett. B \textbf{174}, 45-47 (1986)


\bibitem{Trautner2018}
A.~Trautner,
``Systematic construction of basis invariants in the 2HDM,''
JHEP \textbf{05}, 208 (2019)
[arXiv:1812.02614 [hep-ph]].


\bibitem{Trautner2020}
A.~Trautner,
``On the systematic construction of basis invariants,''
J. Phys. Conf. Ser. \textbf{1586}, no.1, 012005 (2020)
[arXiv:2002.12244 [hep-ph]].




\bibitem{BFHH2007}
S.~Benvenuti, B.~Feng, A.~Hanany and Y.~H.~He,
``Counting BPS Operators in Gauge Theories: Quivers, Syzygies and Plethystics,''
JHEP \textbf{11}, 050 (2007)
[arXiv:hep-th/0608050 [hep-th]].



\bibitem{Molien1897}
T.~Molien,
``{\"U}ber die Invarianten der linearen Substitutionsgruppe,"
Sitzungber. K{\"o}nig. Preuss. Akad. Wiss. (J. Berl. Ber.). 52: 1152–1156


\bibitem{Weyl1926}
H.~Weyl,
``Zur Darstellungstheorie und Invariantenabzählung der projektiven, der Komplex-und der Drehungsgruppe,''
Acta Mathematica 48.3-4 (1926): 255-278.

\bibitem{Casas2001}
J.~A.~Casas and A.~Ibarra,
``Oscillating neutrinos and $\mu \to e, \gamma$,''
Nucl. Phys. B \textbf{618}, 171-204 (2001)
[arXiv:hep-ph/0103065 [hep-ph]].


\bibitem{Pontecorvo1957}
  B.~Pontecorvo,
  ``Mesonium and anti-mesonium,''
  Sov.\ Phys.\ JETP {\bf 6}, 429 (1957)
  [Zh.\ Eksp.\ Teor.\ Fiz.\  {\bf 33}, 549 (1957)].

\bibitem{MNS1962}
  Z.~Maki, M.~Nakagawa and S.~Sakata,
  ``Remarks on the unified model of elementary particles,''
  Prog.\ Theor.\ Phys.\  {\bf 28}, 870 (1962).

\bibitem{Hanany2008}
A.~Hanany, N.~Mekareeya and G.~Torri,
``The Hilbert Series of Adjoint SQCD,''
Nucl. Phys. B \textbf{825}, 52-97 (2010)
[arXiv:0812.2315 [hep-th]].

\bibitem{Haba1998}
N.~Haba, N.~Okamura and M.~Sugiura,
``The Renormalization group analysis of the large lepton flavor mixing and the neutrino mass,''
Prog. Theor. Phys. \textbf{103}, 367-377 (2000)
[arXiv:hep-ph/9810471 [hep-ph]].

\bibitem{Casas1999}
J.~A.~Casas, J.~R.~Espinosa, A.~Ibarra and I.~Navarro,
``Naturalness of nearly degenerate neutrinos,''
Nucl. Phys. B \textbf{556}, 3-22 (1999)
[arXiv:hep-ph/9904395 [hep-ph]].

\bibitem{Esteban:2020cvm}
I.~Esteban, M.~C.~Gonzalez-Garcia, M.~Maltoni, T.~Schwetz and A.~Zhou,
``The fate of hints: updated global analysis of three-flavor neutrino oscillations,''
JHEP \textbf{09}, 178 (2020)
[arXiv:2007.14792 [hep-ph]].


\bibitem{Pilaftsis1997}
A.~Pilaftsis,
``CP violation and baryogenesis due to heavy Majorana neutrinos,''
Phys. Rev. D \textbf{56}, 5431-5451 (1997)
[arXiv:hep-ph/9707235].


\bibitem{Branco2001}
  G.~C.~Branco, T.~Morozumi, B.~M.~Nobre and M.~N.~Rebelo,
  ``A Bridge between CP violation at low-energies and leptogenesis,''
  Nucl.\ Phys.\ B {\bf 617}, 475 (2001)
  [hep-ph/0107164].


\bibitem{CIP2006}
  V.~Cirigliano, G.~Isidori and V.~Porretti,
  ``CP violation and Leptogenesis in models with Minimal Lepton Flavour Violation,''
  Nucl.\ Phys.\ B {\bf 763}, 228 (2007)
  [hep-ph/0607068].



\bibitem{XZ2011}
  Z.~z.~Xing and S.~Zhou,
  ``Neutrinos in particle physics, astronomy and cosmology,''
  Springer-Verlag, Berlin Heidelberg (2011).


\bibitem{Gu2006}
P.~H.~Gu, H.~Zhang and S.~Zhou,
``A Minimal Type II Seesaw Model,''
Phys. Rev. D \textbf{74}, 076002 (2006)
[arXiv:hep-ph/0606302 [hep-ph]].


\end{thebibliography}
\end{document}